\documentclass[aps,preprint,superscriptaddress,nofootinbib]{revtex4-1}

\usepackage{amsmath}
\usepackage{fixmath}
\usepackage{hyperref}
\usepackage{graphicx}
\usepackage{subcaption}
\usepackage{appendix}
\usepackage{xcolor}

\allowdisplaybreaks

\newcommand{\mb}{\mathbold}

\begin{document}

\title{Relaxation operator for quasiparticles in a solid}

\author{Maria Erukhimova}
\affiliation{Institute of Applied Physics, Russian Academy of Sciences}
\author{Yongrui Wang}
\affiliation{Department of Physics and Astronomy, Texas A\&M
	University, College Station, TX, 77843 USA}
\author{Mikhail Tokman}
\affiliation{Institute of Applied Physics, Russian Academy of Sciences}
\author{Alexey Belyanin}
\affiliation{Department of Physics and Astronomy, Texas A\&M
	University, College Station, TX, 77843 USA}

\begin{abstract}
	Popular models of the phenomenological relaxation operators that are widely used in the master equation formalism for open condensed-matter systems have significant flaws ranging from limited applicability to violation of fundamental physical principles. We propose a relatively simple universal model of the relaxation operator which is free from these flaws, has a correct static limit, correct direct-current limit in a uniform electric field, includes both interband and intraband transitions, and is valid for an arbitrary dispersion of quasiparticles in a solid. We use the proposed operator to generalize the Lindhard formula and derive explicit expressions for the relaxation operator for Dirac materials with an unconventional energy spectrum of quasiparticles, such as graphene and Weyl semimetals. We compare the linear susceptibility spectra for graphene obtained with different relaxation models and show that the proposed relaxation operator leads to physically meaningful behavior of the susceptibility at low frequencies whereas the existing models become completely invalid.  
\end{abstract}

\maketitle

\section{Introduction}

The description of open quantum systems is often based on the master equation with a relaxation operator \cite{breuer,blum}
\begin{equation} \label{eq_1_} 
\frac{\partial \hat{\rho}}{\partial t}+\frac{i}{\hbar}\left[\hat{H},\hat{\rho}\right]=\hat{R}\left(\hat{\rho}\right).                                                                     
\end{equation} 
There are many approximations to the form of the relaxation operator that make Eq.~(\ref{eq_1_}) more tractable.  Phenomenological models are particularly popular because of their simplicity. A hybrid approach is also possible, which  combines the microscopic description of relaxation of populations with phenomenological models of relaxation of quantum coherences; see, for example, \cite{wang2015,winzer2013}. In the energy basis the populations and quantum coherences correspond to the diagonal and off-diagonal elements of density matrix, respectively. The choice of phenomenological models was discussed in a number of papers \cite{mermin1970,tokman2013,zhang2014,salmilehto2012,tokman2009,atwal2002}. Here we derive a universal and relatively simple expression for the relaxation operator of quantum coherences in an ensemble of quasiparticles in a solid, which is free from inconsistencies typical for the known models. We use this operator to generalize the Lindhard formula \cite{haug2009quantum} and consider the case of a dissipative 2D system such as graphene as an example.

The simplest phenomenological relaxation operator has the following form in the energy basis \cite{fain1969quantum}:
\begin{equation} \label{eq_2_} 
R_{\alpha \beta}=-{\gamma}_{\alpha \beta}\left[{\rho}_{\alpha \beta}-{\delta}_{\alpha \beta}n^{\left(0\right)}_{\beta}\right],                                                             
\end{equation} 
where $ {\gamma}_{\alpha \beta} $  is the relaxation rate for the transition $ \alpha \leftrightarrow \beta $ and $ n^{\left(0\right)}_{\beta} $ are equilibrium populations. This model corresponds to the well-known replacement $ \omega \Rightarrow \omega +i\gamma $ for the equal constants $ {\gamma}_{\alpha \beta}= \gamma $. 
For the coherences such a relaxation operator is in agreement with the well-known Lindblad form \cite{fain1969quantum,lindblad1975,scully1997quantum}, whereas the diagonal elements according to Eq.~\eqref{eq_2_} relax to the equilibrium state.
Unfortunately, this popular model has serious flaws as described below. 

\subsection{Violation of the continuity equation}

Using expression \eqref{eq_2_} in Eq.~(\ref{eq_1_}) can lead to a number of inconsistencies and mistakes. First of all, it leads to violation of the continuity equation connecting the charge density and the current density in a distributed system, as well as an incorrect stationary perturbation limit \cite{mermin1970,tokman2013}. As a consequence of the violation of the continuity equation, for a bounded isolated system in an alternating external field the relation $ \mb{J}=\frac{\partial}{\partial t}\mb{P} $ between the dipole moment of the system $ \mb{P} $ and the average current $ \mb{J} $ is no longer valid \cite{tokman2013}.  For $ \mathrm{\propto e}^{i\mb{\kappa}\mb{r}-i\omega t} $ processes the violation of the continuity equation leads to violation of the standard relation $ \omega Q_{\mb{\kappa}\omega}=\mb{\kappa}{\mb{j}}_{\mb{\kappa}\omega} $ between the Fourier harmonics of the charge density $ Q_{\mb{\kappa}\omega} $ and current density $ {\mb{j}}_{\mb{\kappa}\omega} $. As a result, if one calculates the conductivity $ \sigma \left(\omega ,\mb{\kappa}\right) $ and polarizability $ \chi \left(\omega ,\mb{\kappa}\right) $ independently, the fundamental relationship between them, namely
\begin{equation} \label{eq_3_} 
\sigma \left(\omega ,\mb{\kappa}\right)=-i\omega \chi \left(\omega ,\mb{\kappa}\right), 
\end{equation} 
turns out to be satisfied only with an accuracy of the order of $ \sim {\gamma}/{\omega} $. Therefore, one has to choose which of these two quantities is more ``correct'' or adequate for a particular situation and calculate it in the framework of the particular microscopic model of the material. In this case Eq.~\eqref{eq_3_} has to be considered as a definition that allows one to find another quantity (see, for example, \cite{tokman2009}). This is hardly acceptable, since there is no universal rule for choosing which response function is ``correct'': $ \sigma $ or $ \chi $. When describing high-frequency or low-dissipation processes in which $ \omega \gg \gamma $, this inconsistency does not lead to significant errors. At the same time, in the region of relatively low frequencies the use of Eq.~\eqref{eq_2_} is highly problematic (see, for example, \cite{tokman2013}). 

Of particular interest in this regard is the description of Coulomb screening in a dissipative system. In this case, Mermin \cite{mermin1970} proposed a modified relaxation operator, which can be represented as follows:
\begin{equation} \label{eq_4_} 
R_{\alpha \beta}=-\gamma \left[{\rho}_{\alpha \beta}-{\delta}_{\alpha \beta}n^{\left(0\right)}_{\beta}-{\eta}^{\left(\mathrm{st}\right)}_{\alpha \beta}\left(\delta \mu \right)\right],                                                             
\end{equation} 
where $ {\eta}^{\left(\mathrm{st}\right)}_{\alpha \beta}\left(\delta \mu \right) $ is a quasistationary perturbation of the equilibrium density matrix, which is linear in perturbation of the chemical potential $ \delta \mu $. For $ {\propto \mathrm{e}}^{i\mb{\kappa}\mb{r}-i\omega t} $ processes, Mermin  \cite{mermin1970} developed the procedure which allows one to find the solution for $ \delta \mu \left(\mb{\kappa},\omega \right) $ which preserves the continuity equation. The latter guarantees that, when the relaxation operator \eqref{eq_4_} is used, the relation \eqref{eq_3_} is satisfied in which the conductivity and polarizability are calculated independently. In Eq.~\eqref{eq_4_} the matrix $ {\eta}^{\left(\mathrm{st}\right)}_{\alpha \beta}\left(\delta \mu \right) $ does not depend on the relaxation constant, since it corresponds to the equilibrium state which the system approaches for a stationary perturbation (i.e, when $ \omega \to  0 $), regardless of the relaxation mechanism. This approach goes back to the paper by Landau \cite{landau1935} on the theory of the dispersion of the magnetic permeability in ferromagnetic media. 

It is important to note that the procedure proposed in \cite{mermin1970} is limited to the simplest case, when the plane waves are considered as basic eigenstates of the unperturbed Hamiltonian, and the energy dispersion of the carriers is parabolic with respect to the quasimomentum $ \mb{k} $, i.e.~it corresponds to the electron current being proportional to the electron quasi-momentum:  $ \mb{j}=-\frac{e}{m}\hbar \mb{k} $, where $ -e $ is an electron charge and $ m $ is a fixed effective mass.

\subsection{The static limit}

The second important test of the relaxation operator model is the behavior of the solution to the master equation (\ref{eq_1_})  in the limit of a static perturbing potential. In this case a \textit{closed system} should reach an equilibrium state in a given external potential. Such a state should not depend on the nature and rate of relaxation, and there is obviously no current in it. As a result, the following requirements appear reasonable: (\textbf{i}) for any $ \mb{\kappa} $ the quantity $ \lim\limits_{\omega \to 0}\mathrm{Re}[\chi \left(\omega ,\mb{\kappa}\right)]=\lim\limits_{\omega \to 0}{\omega}^{-1}\mathrm{Im}[\sigma \left(\omega ,\mb{\kappa}\right)] $ should not depend on the parameters and the model of relaxation, (\textbf{ii}) $ \lim\limits_{\omega \to 0}\omega \mathrm{Im}[\chi \left(\omega ,\mb{\kappa}\right)]=\lim\limits_{\omega \to 0}\mathrm{Re}[\sigma \left(\omega ,\mb{\kappa}\right)]=0 $. However, such a solution cannot describe the situation in which a conductive sample with boundaries which are permeable for carriers is an element of a direct current circuit. In the latter case the limit $ \lim\limits_{ \substack{ \mb{\kappa}\to 0 \\ \omega \to 0}} \omega \mathrm{Im}\chi \left(\omega ,\mb{\kappa}\right)=\lim\limits_{ \substack{ \mb{\kappa}\to 0 \\ \omega \to 0}} \mathrm{Re}\sigma \left(\omega ,\mb{\kappa}\right)=\sigma \left(\gamma \right) $ is nonzero and should correspond to the ohmic conductivity in the uniform constant field, which depends on the relaxation constant $ \gamma $. There is no contradiction with the previous statement, since the element of an electric circuit is obviously not a closed system. A continuous transition from the equilibrium current-free solution to the Ohmic conductivity is possible only within the framework of a problem with boundary conditions.

 The current-free steady state can be obtained by expanding the initial equations in powers of a small parameter which is independent on the relaxation constant, $ \displaystyle \frac{e\delta \mathrm{\Phi}}{\langle W\rangle} $, where $ \delta \mathrm{\Phi} $ is a maximum potential drop and  $ \langle W\rangle $ is a characteristic electron energy. The state with direct current satisfying Ohm's law can be obtained by expanding in powers of another small parameter: $ \displaystyle  \frac{eE}{\gamma \langle p\rangle} $, which includes the relaxation constant $ \gamma $, the characteristic value of the electric field $ E $, and the characteristic momentum $ \langle p\rangle $ of the carriers in the conduction band. Note that under the condition $  \displaystyle \frac{eE}{\gamma \langle p\rangle}\ll 1 $ the initial equilibrium distribution of carriers in the conduction band is weakly perturbed for any ratio $ \displaystyle  \frac{e\delta \mathrm{\Phi}}{\langle W\rangle} $.

 Let us discuss how the above properties relate to the functions $ \sigma \left(\omega ,\mb{\kappa}\right) $ and $ \chi \left(\omega ,\mb{\kappa}\right) $,  obtained as a result of solving the master equation with relaxation operators defined by Eq.~\eqref{eq_2_} and Eq.~\eqref{eq_4_}, respectively, for conduction electrons in a metal or in a bulk semiconductor far from any boundaries.

 For a standard model given by Eq.~\eqref{eq_2_}, $ \lim\limits_{\mb{\kappa}\to 0}\sigma \left(\omega ,\mb{\kappa}\right) $ corresponds to the complex Drude conductivity in a uniform field. In the limit $ \lim\limits_{ \substack{ \mb{\kappa}\to 0 \\  \omega \to 0}} \mathrm{Re}[\sigma \left(\omega ,\mb{\kappa}\right)]=\sigma \left(\gamma \right) $, one obtains the standard Ohmic conductivity in a \textit{uniform and time-independent} field. At the same time, for finite values of $ \mb{\kappa} $ and $ \omega \to 0 $ the expression for the conductivity is incorrect: there is no solution corresponding to the equilibrium current-free state, since $ \lim\limits_{\omega \to 0}\mathrm{Re}[\sigma \left(\omega ,\mb{\kappa}\right)] \neq 0 $. As noted above, for the relaxation model of Eq.~\eqref{eq_2_} the relationship  \eqref{eq_3_} is violated. Therefore, independent calculations of the conductivity and susceptibility $ \chi \left(\omega ,\mb{\kappa}\right) $ lead to different results, but both of them are incorrect:  in the limit $ \omega \to 0 $ and for finite values of $ \mb{\kappa} $ the  expression for $ \lim\limits_{\omega \to 0}\mathrm{Re}[\chi \left(\omega ,\mb{\kappa}\right)]  $ depends on the relaxation parameter $ \gamma $.

 The model based on Eq.~\eqref{eq_4_} preserves  the continuity equation, and the limit $ \omega \to 0 $ for any finite $ \mb{\kappa} $ corresponds to an equilibrium current-free state in a closed system \cite{mermin1970}. In this case, however, it follows from the relations in \cite{mermin1970}  that the limit  $ \mb{\kappa}\to 0 $ leads to $ \lim\limits_{\mb{\kappa}\to 0}\omega \mathrm{Im}[\chi \left(\omega ,\mb{\kappa}\right)]=0 $, both for finite values of $ \omega $ and after taking the subsequent limit $ \omega \to 0 $.

Thus, the model proposed in \cite{mermin1970} provides more adequate description of the screening effects in comparison with the standard approximation \eqref{eq_2_}, but does not describe Ohmic conductivity in a uniform field. In addition, it cannot include interband transitions and is limited to the quadratic energy dispersion of quasiparticles. Besides, the model is rather complicated.

\subsection{The relaxation operator in a real basis} 

For a real Hamiltonian in the absence of a magnetic field \cite{landau2013quantum} one can always choose the basis  eigenfunctions to be real. In this case a much simpler relaxation operator was proposed in \cite{tokman2013}:
\begin{equation} \label{eq_5_} 
R_{\alpha \beta}=-{\gamma}_{\alpha \beta}\left({\rho}_{\alpha \beta}-{\rho}_{\beta \alpha}\right).                                                              
\end{equation} 
Equation \eqref{eq_5_} does not determine relaxation of the diagonal elements of the density matrix; however, if necessary, the relaxation operator for the populations can be added separately: see, for example, \cite{fain1969quantum,tokman2013,zhang2014}.  Note that the diagonal elements (populations) are usually not perturbed in the linear approximation with respect to an external field. 

 Equation \eqref{eq_5_} was obtained in \cite{tokman2013} from the first principles for a system which possesses an electric dipole-allowed transition interacting with a radiation reservoir. In this case, the interaction of the quantum system with the reservoir is described beyond the rotating wave approximation (RWA), i.e. the interaction Hamiltonian includes off-resonant counter-rotating terms.  The master equation beyond the RWA  was studied also in \cite{fleming2010, stokes2012, munro1996}. The relaxation operator \eqref{eq_5_} is not of the Lindblad form \cite{lindblad1975}. Nevertheless, one can show  \cite{tokman2013,munro1996}  that at times exceeding the averaging time corresponding to the Markov approximation, the use of the relaxation operator \eqref{eq_5_} does not violate the condition of positive definiteness of the density matrix.

In the steady-state case, the solution of Eq.~\eqref{eq_1_} with the relaxation operator \eqref{eq_5_} corresponds to an equilibrium state in a static external field, and this equilibrium state does not depend on the relaxation constants. Since Eq.~\eqref{eq_5_}, in contrast to Eq.~\eqref{eq_4_}, does not explicitly depend on the external field (In Eq.~\eqref{eq_4_} the external field defines the value $ \delta \mu $), this result seems paradoxical. The point, however, is that in the real basis the stationary perturbation corresponds to the real values of $ {\rho}_{\alpha \beta} $, so that the relaxation operator \eqref{eq_5_} is zero.

For time-varying fields, the use of the relaxation operator \eqref{eq_5_} ensures that the relation $ \mb{J}=\frac{\partial}{\partial t}\mb{P} $ is satisfied for a bounded isolated system \cite{tokman2013}. For the simplest systems (harmonic oscillator and free particles) placed in a dissipative reservoir and, simultaneously, in a magnetic field, a generalization of the model based on Eq.~\eqref{eq_5_} was developed in \cite{tokman2013,zhang2014}. They proposed an approach based on the transition from energy representation to the coordinate representation, taking into account the requirement of gauge invariance of the observables in an external field with a nonzero vector potential. This condition imposes certain restrictions on the relaxation operator \cite{tokman2013,tokman2009}.

As we see, there is strong motivation to derive the phenomenological relaxation operator that would have the antisymmetric structure like Eq.~\eqref{eq_5_} and would remain applicable to the most general case. In this paper we obtain such a relaxation operator which

\noindent (\textbf{i}) is valid for charged carriers in solids with an arbitrary energy dispersion, in particular the Dirac spectrum;

\noindent (\textbf{ii}) preserves the continuity equation while including both intraband and interband transitions;

\noindent (\textbf{iii}) allows one to obtain both the stationary ``current-free'' regime in equilibrium and the ohmic direct current  regime in the limit of a uniform static field;

\noindent (\textbf{iv}) is significantly simpler than the model proposed in \cite{mermin1970}.

%%%%%%%%%%%%%%%%%%%%%%%%%

In the simplest (intraband) version, the analog of the expression Eq.~\eqref{eq_5_} for $ |\mb{k}\rangle$-states has the form
\begin{equation} \label{eq_6_} 
R_{\mb{k}\mb{q}}=-{\gamma}_{\mb{kq}}\left({\rho}_{\mb{kq}}-{\rho}_{-\mb{q}-\mb{k}}\right).                                                          
\end{equation} 
This expression has a simple interpretation. Suppose that the transition between the states $ |\mb{k}\rangle \leftrightarrow |\mb{q}\rangle $ is accompanied by some excitation of the ``reservoir''. This excitation must have quasimomentum $ \mb{p} $, such that $ \hbar \mb{k} = \hbar\mb{q}+\mb{p} $. However, in this case, the  conservation of momentum is also valid for the transition $ |{-}\mb{q}\rangle \leftrightarrow |{-}\mb{k}\rangle $, since the relation $ -\hbar \mb{q} =- \hbar \mb{k}+\mb{p} $ is true. Thus, the reservoir modes inevitably ``couple'' the transitions $ |\mb{k}\rangle \leftrightarrow |\mb{q}\rangle $ and $ |{-}\mb{q}\rangle \leftrightarrow |{-}\mb{k}\rangle $, which is reflected in the expression \eqref{eq_6_}.

The structure of the paper is as follows. Section \ref{sec_relaxation} establishes general requirements for the structure of a relaxation operator, following from the conservation of the number of particles. An operator for an ensemble of quasiparticles in a solid is constructed, which ensures the continuity equation to be satisfied exactly or after averaging over the lattice period; see Eqs.~\eqref{eq_25_} below. It turns out that the derivation of such an operator is significantly simplified for systems that are symmetric with respect to time reversal (TRS-systems). Here we mean the corresponding property of the isolated system without taking into account its relatively weak interaction with a dissipative reservoir. In section \ref{sec_lindhard}, the Lindhard formula is obtained for a dissipative 2D system using the correct (in the above sense) relaxation operator. In section \ref{sec_comparison} we compare our results applied to graphene with the results of using the standard model given by Eq.~\eqref{eq_2_}. In  Appendix \ref{sec_appendix_TRS} one property of the TRS systems is established which is important for derivation of the relaxation operator. Appendix \ref{appendix_graphene} and \ref{appendix_WSM} describe the the procedure for deriving a phenomenological relaxation operator in graphene and Weyl semimetals with broken time-reversal symmetry. In Appendix \ref{sec_appendix_homogeneous} the properties of the linear susceptibility in the limit of a uniform high-frequency field are considered. Appendix E contains the derivation of the susceptibility of monolayer graphene in the limit of $ \mb{\kappa} = 0 $.

\section{Relaxation operator preserving the continuity equation} 
\label{sec_relaxation}

\subsection{Basic relationships}

First, we write the Hamiltonian of a nonrelativistic electron in a periodic potential in a fairly general form, 
\[\hat{H}={\hat{H}}_0\left(\mb{r},\hat{\mb{p}}, \hat{\mb{s}}\right),\] 
where the dependence of the Hamiltonian on the coordinate $ \mb{r} $ is periodic, $ \hat{\mb{p}}\mb = -i\hbar \mb{\mathrm{\nabla}} $ is a momentum operator, $ \hat{\mb{s}}=\frac{1}{2}\left({\mb{x}}_0{\hat{\sigma}}_x+{\mb{y}}_0{\hat{\sigma}}_y+{\mb{z}}_0{\hat{\sigma}}_z\right) $ is a spin operator, and  $ {\hat{\sigma}}_{x,y,z} $ are Pauli matrices. The dependence on the spin operator may be, e.g., due to spin-orbit coupling.

 If there are perturbing fields defined by electrodynamic potentials $ \varphi \left(\mb{r},t\right) $ and $ \mb{A}\left(\mb{r},t\right) $, the operator $ \hat{H} $ can be obtained from the unperturbed Hamiltonian $ {\hat{H}}_0\left(\hat{\mb{p}}\right) $ as \cite{landau2013quantum}
\begin{equation} \label{eq_7_} 
\hat{H}={\hat{H}}_0\left(\hat{\mb{p}}\Rightarrow \hat{\mb{p}}+\frac{e}{c}\mb{A}\right)-e\varphi .                                                             
\end{equation} 

For simplicity, we do not consider here the spin-dependent components of the perturbation operator, such as the energy of a spin magnetic moment in a magnetic field, $ {\hat{V}}_B=-{\mu}_B\left[\hat{\mb{s}}\cdot \left(\mb{\mathrm{\nabla}} \times \mb{A}\right)\right] $, where  $ {\mu}_B $ is Bohr's magneton, or the spin-orbit coupling term in the perturbing field, $ {\hat{V}}_{s-o}=-\frac{e\hbar}{2m^2c^2}\left[\left(\frac{1}{c}\dot{\mb{A}}+\mb{\mathrm{\nabla}}\varphi \right)\times \hat{\mb{p}}\right]\cdot \hat{\mb{s}} $ \cite{landau2013quantum,gantmakher1987carrier,berestetskii1982quantum}.

Consider the energy basis given by the stationary solution of the Schr\"{o}dinger equation:
\[{\hat{H}}_0{\mathrm{\Psi}}_{\alpha}\left(\mb{r},s\right)={E_{\alpha}\mathrm{\Psi}}_{\alpha}\left(\mb{r},s\right),\] 
where $ {\mathrm{\Psi}}_{\alpha}\left(\mb{r},s\right) $ are eigenfunctions of the unperturbed Hamiltonian $ {\hat{H}}_0 $, $ \alpha $ is an index or a set of indices indicating a stationary state of the Hamiltonian taking into account spin orientation, and the spin coordinate $ s $ takes the values 1 and 2 denoting spinor components 
 $ \left( \begin{array}{c}
{\mathrm{\Psi}}_{\alpha}\left(\mb{r},1\right)  \\
{\mathrm{\Psi}}_{\alpha}\left(\mb{r},2\right) \end{array}
\right) $, 
which define the probability of spin projection on the quantization axis to be equal to $ \frac{1}{2} $ and  $ -\frac{1}{2} $.

The observed current density $ \mb{j}\left(\mb{r}\right) $ and free carrier density $ n\left(\mb{r}\right) $ can be expressed through the elements of the density matrix in the given basis $ {\rho}_{\alpha \beta} $ as follows (see, for example, \cite{wang2016,kutayiah2018}), 
\begin{equation} \label{eq_8_} 
n\left(\mb{r}\right)=\sum_{\alpha \beta}{\sum^2_{s=1}{\left[{\mathrm{\Psi}}^\ast_{\beta}\left(\mb{r},s\right){\mathrm{\Psi}}_{\alpha}\left(\mb{r},s\right)\right]}{\rho}_{\alpha \beta}},                                                             
\end{equation} 
\begin{equation} \label{eq_9_} 
\mb{j}\left(\mb{r}\right)=-\frac{e}{2}\sum_{\alpha \beta}{\sum^2_{s=1}{\left\{{\mathrm{\Psi}}^\ast_{\beta}\left(\mb{r},s\right)\left[\hat{\mb{v}}{\mathrm{\Psi}}_{\alpha}\left(\mb{r},s\right)\right]+\left[{\hat{\mb{v}}}^\ast{\mathrm{\Psi}}^\ast_{\beta}\left(\mb{r},s\right)\right]{\mathrm{\Psi}}_{\alpha}\left(\mb{r},s\right)\right\}}{\rho}_{\alpha \beta}},          
\end{equation} 
where $ \hat{\mb{v}}=\frac{i}{\hbar}\left[\hat{H},\mb{r}\right]=\frac{1}{m}\left(\hat{\mb{p}}+\frac{e}{c}\mb{A}\right) $ is a velocity operator.

Since our goal in this section is to include the limitations imposed by the conservation of the particle number, we neglected the vortex spin current in Eq.~\eqref{eq_9_},
\[{\mb{j}}_S\left(\mb{r}\right)={\mu}_Bc\mb{\mathrm{\nabla}} \times \sum_{\alpha \beta}{\sum^2_{s=1}{\left[{\mathrm{\Psi}}^\ast_{\beta}\left(\mb{r},s\right)\hat{\mb{s}}{\mathrm{\Psi}}_{\alpha}\left(\mb{r},s\right)\right]{\rho}_{\alpha \beta}}},\] 
because it does not affect the evolution of carrier density.

 For a spin-independent Hamiltonian the space coordinates and spin are separated. The simplest example is  
\[{\hat{H}}_0=-eU\left(\mb{r}\right)+\frac{1}{2m}{\hat{\mb{p}}}^2,\] 
where $ U\left(\mb{r}\right) $ is a periodic lattice potential. In this case, the summation in Eqs.~\eqref{eq_8_}, \eqref{eq_9_} over two equal spin states gives only the degeneracy factor $ g=2 $ in the final expressions:
\begin{equation} \label{eq_10_} 
n\left(\mb{r}\right)=g\sum_{\alpha \beta}{\left[{\mathrm{\Psi}}^\ast_{\beta}\left(\mb{r}\right){\mathrm{\Psi}}_{\alpha}\left(\mb{r}\right)\right]}{\rho}_{\alpha \beta},                                                               
\end{equation} 
\begin{equation} \label{eq_11_} 
\mb{j}\left(\mb{r}\right) =- g\frac{e}{2}\sum_{\alpha \beta}{\left\{{\mathrm{\Psi}}^\ast_{\beta}\left(\mb{r}\right)\left[\hat{\mb{v}}{\mathrm{\Psi}}_{\alpha}\left(\mb{r}\right)\right]+\left[{\hat{\mb{v}}}^\ast{\mathrm{\Psi}}^\ast_{\beta}\left(\mb{r}\right)\right]{\mathrm{\Psi}}_{\alpha}\left(\mb{r}\right)\right\}{\rho}_{\alpha \beta}},                     
\end{equation} 
where the functions $ {\mathrm{\Psi}}_{\alpha}\left(\mb{r}\right) $ are scalar (not the spinors). For brevity, we will use the term ``spinless'' particles. For Fourier harmonics $ n_{\mb{\kappa}}=\frac{1}{{\left(2\pi \right)}^{\varsigma}}\int{n\left(\mb{r}\right){\mathrm{e}}^{-i\mb{\kappa}\mb{r}}d^{\varsigma}r} $ and $ {\mb{j}}_{\mb{\kappa}}=\frac{1}{{\left(2\pi \right)}^{\varsigma}}\int{\mb{j}\left(\mb{r}\right){\mathrm{e}}^{-i\mb{\kappa}\mb{r}}d^{\varsigma}r} $ it follows from Eqs.~\eqref{eq_10_} and \eqref{eq_11_} that 
\begin{equation} \label{eq_12_} 
n_{\mb{\kappa}}=\frac{g}{{\left(2\pi \right)}^{\varsigma}}\sum_{\alpha \beta}{{\left({\mathrm{e}}^{-i\mb{\kappa}\mb{r}}\right)}_{\beta \alpha}}{\rho}_{\alpha \beta},   {\mb{j}}_{\mb{\kappa}}=-\frac{g}{{\left(2\pi \right)}^{\varsigma}}\frac{e}{2}\sum_{\alpha \beta}{{\left({\mathrm{e}}^{-i\mb{\kappa}\mb{r}}\cdot \hat{\mb{v}}+\hat{\mb{v}}{\cdot \mathrm{e}}^{-i\mb{\kappa}\mb{r}}\right)}_{\beta \alpha}{\rho}_{\alpha \beta}},              
\end{equation} 
where $ \varsigma $ is the system dimension.

The evolution of the density matrix is described by the von Neumann equation  
\begin{equation} \label{eq_13_} 
\frac{\partial {\rho}_{\alpha \beta}}{\partial t}+\frac{i}{\hbar}\sum_{\gamma}{\left(H_{\alpha \gamma}{\rho}_{\gamma \beta}-{\rho}_{\alpha \gamma}H_{\gamma \beta}\right)}=0.                                                  
\end{equation} 
By applying the summation operation
\[\sum_{\alpha \beta}{\sum^2_{s=1}{{\mathrm{\Psi}}^\ast_{\beta}{\left(\mb{r},s\right)\mathrm{\Psi}}_{\alpha}\left(\mb{r},s\right)\left(\cdots \right)}}\] 
to Eq.~\eqref{eq_13_} and taking into account Eqs.~\eqref{eq_8_}, \eqref{eq_9_}, we arrive at the continuity equation 
\begin{equation} \label{eq_14_} 
\frac{\partial n}{\partial t}+\frac{\mb{\mathrm{\nabla}}\cdot \mb{j}}{-e}=0.                                                                          
\end{equation}

\subsection{Correct phenomenological relaxation operator}

The correct relaxation operator in Eq.~\eqref{eq_1_} must not violate the continuity equation, i.e. the conservation of the number of particles given by  Eq.~\eqref{eq_14_}. Taking into account Eqs.~\eqref{eq_8_} and \eqref{eq_9_}, the particle conservation law is satisfied under the condition 
\begin{equation} \label{eq_15_} 
\sum_{\alpha \beta} \sum^2_{s=1}{{\mathrm{\Psi}}^\ast_{\beta}\left(\mb{r},s\right){\mathrm{\Psi}}_{\alpha}\left(\mb{r},s\right)R_{\alpha \beta}} = 0,                                                
\end{equation} 
or, for ``spinless'' particles, 
\begin{equation} \label{eq_16_} 
\sum_{\alpha \beta} {\mathrm{\Psi}}^\ast_{\beta}\left(\mb{r}\right){\mathrm{\Psi}}_{\alpha}\left(\mb{r}\right)R_{\alpha \beta} = 0.                                                             
\end{equation} 

As noted in the Introduction, the diagonal elements of the density matrix are usually not perturbed in the linear approximation with respect to the Hamiltonian of interaction with an external field. Therefore, when calculating the linear response of the medium it is sufficient to take into account only the off-diagonal elements of the relaxation operator $ R_{\alpha \beta} $ in the sum \eqref{eq_15_}. 
%In addition, the electron density averaged over interatomic distances in a condensed medium does not depend on the diagonal elements of the density matrix. In this sense, neglecting the diagonal elements in the sum \eqref{eq_15_} is correct also outside the limits of the linear theory. At the same time, the diagonal elements of the relaxation operator must satisfy the condition of the population balance: $ \sum\limits_{\alpha} R_{\alpha \alpha} = 0 $;  an example of the corresponding equations is given, for example, in \cite{fain1969quantum}.

In the basis of real wave functions of ``spinless'' particles, which can always be chosen for a system without a magnetic field \cite{landau2013quantum}, the relaxation operator of the form Eq.~\eqref{eq_5_} ensures that Eq.~\eqref{eq_16_} is satisfied. Such a basis is ``natural'' for a discrete nondegenerate spectrum describing a finite motion.

For particles in a system with translational symmetry (in free space or in a periodic lattice field), the most ``natural'' basis is a set of complex wave functions that are eigenfunctions of the momentum or quasi-momentum operator. When using such a basis, the procedure for constructing the desired relaxation operator similar to Eq.~\eqref{eq_5_} becomes more complicated. The additional complication is caused by spin-dependence of the Hamiltonian. However, as will be shown below, for TRS systems the corresponding procedure is not too cumbersome. It relies on an assumption that the relaxation rate of quantum coherence for a given transition depends only on its energy.

Now let's use the time-reversal symmetry of the system. The operation of reversal in time $ \hat{T} $ as applied to a scalar energy eigenfunction is just an operation of complex conjugation: 
\[\hat{T}{\mathrm{\Psi}}_{\alpha}\left(\mb{r}\right)={\mathrm{\Psi}}^\ast_{\alpha}\left(\mb{r}\right).\] 
When applied to the spinor $ \left( \begin{array}{c}
{\mathrm{\Psi}}_{\alpha}\left(\mb{r},1\right) \\ 
 {\mathrm{\Psi}}_{\alpha}\left(\mb{r},2\right) \end{array}
\right) $ this operation takes the form \cite{landau2013quantum}
\begin{equation} \label{eq_17_} 
\hat{T}\left( \begin{array}{c}
{\mathrm{\Psi}}_{\alpha}\left(\mb{r},1\right) \\ 
{\mathrm{\Psi}}_{\alpha}\left(\mb{r},2\right) \end{array}
\right)=i{\hat{\sigma}}_y\left( \begin{array}{c}
{\mathrm{\Psi}}^\ast_{\alpha}\left(\mb{r},1\right) \\ 
{\mathrm{\Psi}}^\ast_{\alpha}\left(\mb{r},2\right) \end{array}
\right)=\left( \begin{array}{c}
{\mathrm{\Psi}}^\ast_{\alpha}\left(\mb{r},2\right) \\ 
-{\mathrm{\Psi}}^\ast_{\alpha}\left(\mb{r},1\right) \end{array}
\right).                                     
\end{equation}

The Hamiltonian of a TRS-system commutes with the operator $ \hat{T} $. This property is inherent in closed systems, and closure implies, among other things, the absence of an external magnetic field. The Hamiltonian of such systems satisfies the condition   $ \hat{H}\left(\hat{\mb{p}},\hat{\mb{s}}\right)=\hat{H}\left(-\hat{\mb{p}},-\hat{\mb{s}}\right) $. The presence of an external dc electric field does not affect the symmetry with respect to time reversal \cite{landau2013quantum}.  For ``spinless'' particles, such commutativity is equivalent to the Hamiltonian being real in the  $r$-representation. 

To construct a correct relaxation operator we make use of the fact that for every basis state $|\alpha \rangle$ the state $\hat{T}|\alpha \rangle$ coincides with the same or another basis state, up to a constant phase: 
\begin{equation} \label{eq_19_} 
|{\alpha}^\prime \rangle ={\mathrm{e}}^{i{\varphi}_{\alpha}}\hat{T}|\alpha \rangle. 
%\;  |\alpha \rangle ={\mathrm{e}}^{i{\varphi}_{{\alpha}^\prime}}\hat{T}|{\alpha}^\prime \rangle .  
\end{equation}
Such a basis always exists since it can consist of eigenfunctions of operator $\hat{T}$ which commutes with the Hamiltonian, $\left[\hat{H},\hat{T}\right]=0$. In the latter case $|{\alpha}^\prime \rangle =|\alpha \rangle$. For the degenerate energy levels there is a freedom in the choice of basis states, and the states $|{\alpha}^\prime \rangle$ and $|\alpha \rangle$ can be different. However, the condition \eqref{eq_19_} should be satisfied, since in the general case the state $\hat{T}|\alpha \rangle$ can be equal to the linear combination of basis states corresponding to the same energy level.

%As a result, state $ |{\alpha}^\prime \rangle $ is the eigenstate of the Hamiltonian with the same energy value as state $ |{\alpha} \rangle $: $ E_{\alpha}=E_{{\alpha}^\prime} $. For linearly independent states corresponding to the degenerate energy level $ E_{\alpha} $, we choose the basis functions so that there is a one-to-one correspondence between the states $ |\alpha \rangle $ and $ |{\alpha}^\prime \rangle $ defined by Eq.~\eqref{eq_19_}. This relationship, as one can see, is set by the operation of time reversal with proper adjustment of the phase multiplier. 

Since $ {\hat{T}}^2=\pm 1 $ where the upper sign is for a scalar state function and the lower sign is for a spinor, respectively, we obtain that  Eq.~\eqref{eq_19_} is reciprocal:
\begin{equation} \label{eq_19_1} 
%|{\alpha}^\prime \rangle ={\mathrm{e}}^{i{\varphi}_{\alpha}}\hat{T}|\alpha \rangle 
|\alpha \rangle ={\mathrm{e}}^{i{\varphi}_{{\alpha}^\prime}}\hat{T}|{\alpha}^\prime \rangle                                            
\end{equation}
and for a ``spinless'' particle $ {\varphi}_{{\alpha}^\prime}={\varphi}_{\alpha} $, while for a particle described by the spinor $ {\varphi}_{{\alpha}^\prime}={\varphi}_{\alpha}+\pi $.
In both cases for any pair of states $|\alpha \rangle$ and $|\beta \rangle$ 
\begin{equation} \label{eq_20_} 
{\varphi}_{{\beta}^\prime}-{\varphi}_{{\alpha}^\prime}={\varphi}_{\beta}-{\varphi}_{\alpha}.                                                             \end{equation} 
Another useful relation follows from Eqs.~\eqref{eq_17_} and \eqref{eq_19_}: \begin{equation} \label{eq_21_} 
\sum^2_{s=1}{\left[{\mathrm{\Psi}}^\ast_{\beta}\left(\mb{r},s\right){\mathrm{\Psi}}_{\alpha}\left(\mb{r},s\right)-{{\mathrm{e}}^{i\left({{\varphi}_{\alpha}-\varphi}_{\beta}\right)}\mathrm{\Psi}}^\ast_{{\alpha}^\prime}\left(\mb{r},s\right){\mathrm{\Psi}}_{{\beta}^\prime}\left(\mb{r},s\right)\right]}=0.              \end{equation}

Using Eq.~\eqref{eq_19_} we construct the relaxation operator in the following way:
\begin{equation} \label{eq_18_} 
R_{\alpha \beta}=-{\gamma}_{\alpha \beta}\left({\rho}_{\alpha \beta}-{{\mathrm{e}}^{i\left({\varphi}_{\beta}-{\varphi}_{\alpha}\right)}\rho}_{{\beta}^\prime {\alpha}^\prime}\right).                      
\end{equation}
Calculating  the sum of the off-diagonal elements in Eq.~\eqref{eq_15_} with the relaxation operator \eqref{eq_18_}, using Eqs.~\eqref{eq_20_} and  \eqref{eq_21_}, and rearranging the summation indices, we confirm that Eq.~\eqref{eq_15_} is satisfied, which, as stated earlier, is the criterion ensuring the continuity equation in the system.

Consider the following examples: (\textbf{i}) free particles, (\textbf{ii}) ``spinless'' particles in a periodic lattice and (\textbf{iii}) spin-dependent system in a periodic lattice. 

\noindent (\textbf{i}) In this case we have $ |\alpha \rangle =|\mb{k}\rangle $, where a set of vectors $ \mb{k} $ is given by periodic boundary conditions. In the coordinate representation the wave function has the form $ {\mathrm{\Psi}}_{\mb{k}}\left(\mb{r}\right)={\mathrm{e}}^{i\mb{kr}} $, $ {{\mathrm{\Psi}}_{{\mb{k}}^\prime}\left(\mb{r}\right) = \mathrm{\Psi}}^\ast_{\mb{k}}={\mathrm{\Psi}}_{-\mb{k}} $,  
so that the corresponding phases ${\varphi}_{\alpha}$ and ${\varphi}_{{\alpha}^\prime}$ in Eqs.~\eqref{eq_19_} and \eqref{eq_19_1}  are equal to zero. 
Then Eq.~\eqref{eq_18_} gives  
\[R_{\mb{k}\mb{q}}=-{\gamma}_{\mb{kq}}\left({\rho}_{\mb{kq}}-{\rho}_{-\mb{q}-\mb{k}}\right).\] 
A simple interpretation of this expression is discussed in the Introduction. 

\noindent (\textbf{ii}) In this case we have $ |\alpha \rangle =|c,\mb{k}\rangle $, where $c$ is a band index. In the coordinate representation the wave function has a  form of the Bloch function: $ {\mathrm{\Psi}}_{c\mb{k}}\left(\mb{r}\right)={\psi}_{c\mb{k}}\left(\mb{r}\right){\mathrm{e}}^{i\mb{kr}} $, where $ {\psi}_{c\mb{k}}\left(\mb{r}\right) $  is a periodic function with the lattice period, or a sum of periodic functions (when several sublattices exists). For a real Hamiltonian in the $\mb{r}$-representation, the dependence of the electron energy on the quasimomentum in a given band is a symmetric function: $ E_{c\mb{k}}=E_{c-\mb{k}} $, $ {{\mathrm{\Psi}}_{c-\mb{k}} = {\mathrm{e}}^{i{\varphi}_{c\mb{k}}}\mathrm{\Psi}}^\ast_{c\mb{k}} $. Thus, each energy state is at least four times degenerate: twice in quasi-momentum and twice in spin. It follows from Eq.~\eqref{eq_18_} that
\begin{equation} \label{eq_22_} 
R_{c\mb{k}d\mb{q}}=-{\gamma}_{c\mb{k}d\mb{q}}\left({\rho}_{c\mb{k}d\mb{q}}-{\mathrm{e}}^{i\left({\varphi}_{d\mb{q}}-{\varphi}_{c\mb{k}}\right)}{\rho}_{d\, -\mb{q}\, c\, -\mb{k}}\right).                                    
\end{equation} 
(\textbf{iii}) For a spin-dependent periodic Hamiltonian, spin degeneracy can be lifted. In this case we have $ {\mathrm{\Psi}}_{c\mb{k}}\left(\mb{r},s\right)={\psi}_{c\mb{k}}\left(\mb{r},s\right){\mathrm{e}}^{i\mb{kr}} $, where periodic functions are the components of the spinor $ {\psi}_{c\mb{k}}\left(\mb{r},1\right) $ and $ {\psi}_{c\mb{k}}\left(\mb{r},2\right) $. The energies of electron states with opposite quasi-momenta and, simultaneously, with opposite average values of spin projections onto a given axis, turn out to be equal. Such states are connected by the operation of time reversal. Thus, we have $ E_{c\mb{k}}=E_{c^\prime -\mb{k}} $, where:
\[\left( \begin{array}{c}
{\psi}_{c^\prime -\mb{k}}\left(\mb{r},1\right) \\ 
{\psi}_{c^\prime -\mb{k}}\left(\mb{r},2\right) \end{array}
\right){\mathrm{e}}^{-i\mb{kr}}={\mathrm{e}}^{i{\varphi}_{c\mb{k}}}{\left[\left( \begin{array}{c}
{\psi}_{c\mb{k}}\left(\mb{r},2\right) \\ 
-{\psi}_{c\mb{k}}\left(\mb{r},1\right) \end{array}
\right){\mathrm{e}}^{i\mb{kr}}\right]}^\ast.\] 
To avoid any confusion, we emphasize that the band index $c$ numbers all states with different energies for a given $ \mb{k} $. The bands created by  spin splitting correspond to different values of the index $c$. When $ \mb{k}=0 $, such bands intersect or touch, so that the degeneracy is restored. Indeed, if the state with $ \mb{k}=0 $ and energy $ E_{c\mb{k}=0} $ is described by the spinor $ \left( \begin{array}{c}
{\psi}_c\left(\mb{r},1\right) \\ 
{\psi}_c\left(\mb{r},2\right) \end{array}
\right) $, then the same energy level corresponds to the state $ \hat{T}\left( \begin{array}{c}
{\psi}_c\left(\mb{r},1\right) \\ 
{\psi}_c\left(\mb{r},2\right) \end{array}
\right)=\left( \begin{array}{c}
{\psi}^\ast_c\left(\mb{r},2\right) \\ 
-{\psi}^\ast_c\left(\mb{r},1\right) \end{array}
\right) $, linearly independent with the first state. Therefore, the state with energy $ E_{c\mb{k}=0} $ is degenerate. The choice of band numbering at the crossing or contact point is a matter of convention. The relaxation operator \eqref{eq_18_} again has the form of Eq.~\eqref{eq_22_} if we consider the band index to be conserved when the sign of the quasi-momentum changes, i.e. choose band numbering so that $ E_{c\mb{k}}=E_{c-\mb{k}} $.

\subsection{Averaged relaxation operator}

The relaxation operator satisfying the continuity equation for carriers in a solid couples coherences at the transitions that are symmetric in the quasimomentum space with respect to the point $ \mb{k}=0 $. It often makes sense to restrict ourselves to the carrier states in a relatively small vicinity of a certain point $ {\mb{k}}_{0} $ of the Brillouin zone. Suppose that such a point is an extremum, in the vicinity of which the carrier dispersion can be considered symmetric: $ E_c\left({\mb{k}-\mb{k}}_{0}\right)=E_c\left({\mb{k}}_{0}-\mb{k}\right) $ (for example, the vicinity of the Dirac points \cite{vafek2014,orlita2014}). Whenever we calculate any observable quantities including only the states within a small part of the Brillouin zone, we in fact determine their values  averaged over a scale much longer than the lattice period $a$. In this case it makes sense to require that the continuity equation be satisfied also ``on average'' over the same scales. It turns out that a relatively simple relaxation operator preserving the number of particles ``on average'' can be constructed without requiring TRS of the system.

%Within the averaged description, the diagonal elements of the relaxation operator do not affect the  continuity equation in any way, since the redistribution of populations does not perturb the averaged carrier density. 
The corresponding averaging implies a rather narrow interval of quasi-momenta $ \delta k $ satisfying 
\begin{equation} \label{eq_23_} 
\delta k a \ll 1.                                                                                     
\end{equation} 
Let us average Eq.~\eqref{eq_15_} over the lattice period, assuming that inequality \eqref{eq_23_} is satisfied:
\begin{equation} \label{eq_24_} 
 \sum_{cd\mb{k}\mb{q}}{\sum^2_{s=1}{\overline{{\psi}^\ast_{d\mb{q}}{\left(\mb{r},s\right)\psi}_{c\mb{k}}\left(\mb{r},s\right)}}{\mathrm{e}}^{i\left(\mb{k}-\mb{q}\right)\mb{r}}R_{c\mb{k}d\mb{q}}}=0,                                
\end{equation} 
where the bar denotes the corresponding averaging, and the quasimomentum is counted from the point $ {\mb{k}=\mb{k}}_{0} $. 

The diagonal terms $c\mb{k}= d\mb{q}$ in Eq.~\eqref{eq_24_} give zero contribution to the sum, since the diagonal terms under the averaging bar are equal to one (normalization condition), the exponential terms are also equal to one, and the conservation of the total number of particles in the system requires that  $ \sum\limits_{c\mb{k}} R_{c\mb{k}c\mb{k}} = 0 $. The remaining sum of the off-diagonal terms determines the coordinate-dependent part of the particle density which has nonzero spatial harmonics and contributes to the charge continuity equation. Therefore, only the off-diagonal components of the relaxation operator determine whether the continuity equation is preserved and Eq.~\eqref{eq_3_} is satisfied. In the averaged description this is true even beyond the linear response theory. In the non-averaged description, the diagonal terms can be also coordinate-dependent and contribute to the continuity equation, as is clear from Eq.~\eqref{eq_15_}. However, within the linear response theory the diagonal terms are not perturbed by the field and only the off-diagonal elements of the relaxation operator need to be considered.

We will seek the matrix of the relaxation operator in a form close to Eq.~\eqref{eq_22_}, replacing factor $ {\mathrm{e}}^{i\left({\varphi}_{d\mb{q}}-{\varphi}_{c\mb{k}}\right)} $ with some matrix element $ G_{c\mb{k}d\mb{q}} $ which we need to determine:
\begin{equation} \label{eq_25_} 
R_{c\mb{k}d\mb{q}}=-{\gamma}_{c\mb{k}d\mb{q}}\left({\rho}_{c\mb{k}d\mb{q}}-G_{c\mb{k}d\mb{q}}{\rho}_{d\,-\mb{q}\,c\,-\mb{k}}\right).                                             
\end{equation} 
Substituting Eq.~\eqref{eq_25_} in Eq.~\eqref{eq_24_}, we get
\[\sum_{c\mb{k}d\mb{q}}{{{\gamma}_{c\mb{k}d\mb{q}}\mathrm{e}}^{i\left(\mb{k}-\mb{q}\right)\mb{r}}\left[\sum^2_{s=1}{\overline{{\psi}^\ast_{d\mb{q}}{\left(\mb{r},s\right)\psi}_{c\mb{k}}\left(\mb{r},s\right)}}\left({\rho}_{c\mb{k}d\mb{q}}-G_{c\mb{k}d\mb{q}}{\rho}_{d\,-\mb{q}\,c\,-\mb{k}}\right)\right]}=0,\] 
where after rearranging summation indices,
\begin{equation} \label{eq_26_} 
\sum_{c\mb{k}d\mb{q}}{{{\gamma}_{c\mb{k}d\mb{q}}\mathrm{e}}^{i\left(\mb{k}-\mb{q}\right)\mb{r}}\sum^2_{s=1}{\left[\overline{{\psi}^\ast_{d\mb{q}}{\left(\mb{r},s\right)\psi}_{c\mb{k}}\left(\mb{r},s\right)}-\overline{{\psi}^\ast_{c-\mb{k}}{\left(\mb{r},s\right)\psi}_{d-\mb{q}}\left(\mb{r},s\right)}G_{d\,-\mb{q}\,c\,-\mb{k}}\right]{\rho}_{c\mb{k}d\mb{q}}}}=0.    
\end{equation} 
As a result, we obtain the following expression for the matrix elements $ G_{c\mb{k}d\mb{q}} $, 
\begin{equation} \label{eq_27_} 
G_{c\mb{k}d\mb{q}}=\frac{1}{G_{d\,-\mb{q}\,c\,-\mb{k}}}=\frac{\sum^2_{s=1}{\overline{{\psi}^\ast_{c-\mb{k}}{\left(\mb{r},s\right)\psi}_{d-\mb{q}}\left(\mb{r},s\right)}}}{\sum^2_{s=1}{\overline{{\psi}^\ast_{d\mb{q}}{\left(\mb{r},s\right)\psi}_{c\mb{k}}\left(\mb{r},s\right)}}} 
\end{equation} 
or, for ``spinless'' particles, 
\begin{equation} \label{eq_28_} 
G_{c\mb{k}d\mb{q}}=\frac{1}{G_{d\,-\mb{q}\,c\,-\mb{k}}}=\frac{\overline{{\psi}^\ast_{c-\mb{k}}{\left(\mb{r}\right)\psi}_{d-\mb{q}}\left(\mb{r}\right)}}{\overline{{\psi}^\ast_{d\mb{q}}{\left(\mb{r}\right)\psi}_{c\mb{k}}\left(\mb{r}\right)}}.                                                        
\end{equation} 

A different approach to describe quasiparticles in a crystal is to work with wave functions averaged over the lattice period. The truncated model Hamiltonian that defines such states can be reconstructed using the matrix $ \hat{E}\left(\mb{k}\right) $, whose eigenvalues define the energy bands and the carrier energy dispersion in each band (see, for example, \cite{gantmakher1987carrier}). This matrix can be calculated in various approximations,  including the magnetic and/or spin effects  ``hidden'' in the form of the matrix $ \hat{E}\left(\mb{k}\right) $.

In the region of \textit{k}-space corresponding to Eq.~\eqref{eq_23_} and in the absence of perturbing external fields, the averaged Hamiltonian has the form:
\begin{equation} \label{eq_29_} 
{\hat{H}}_0=\frac{1}{\hbar}{\left(\frac{\partial}{\partial \mb{k}}\hat{E}\right)}_{\mb{k}=0}\cdot \hat{\mb{p}}+\frac{1}{2{\hbar}^2}\sum_{ij}{{\left(\frac{{\partial}^2\hat{E}}{\partial k_i\partial k_j}\right)}_{\mb{k}=0}}\cdot {\hat{p}}_i{\hat{p}}_j,                                       
\end{equation} 
where $ i,j=x,y,z $. The Hamiltonian $ {\hat{H}}_0\left(\hat{\mb{p}}\right) $ is generally an $ N\times N $ matrix which defines a basis of states in the form of \textit{N}-component vectors corresponding to the energies $ E_{c\mb{k}} $:
\begin{equation} \label{eq_30_} 
{\mb{\mathrm{U}}}_{c\mb{k}}\left(\mb{r}\right)\equiv {\mb{u}}_{c\mb{k}}{\mathrm{e}}^{i\mb{kr}}, \, {\mb{u}}_{c\mb{k}}=\left( \begin{array}{c}
u^{\left(1\right)}_{c\mb{k}} \\ 
\vdots  \\ 
u^{\left(N\right)}_{c\mb{k}} \end{array}
\right).                                                              
\end{equation} 
The elements of the vector $ {\mb{u}}_{c\mb{k}} $ in Eq.~\eqref{eq_30_} are the coefficients of the expansion of the Bloch function over orthogonal periodic functions or over orthogonal periodic two-component functions (spinors). The scalar product 
\[\left({\mb{u}}^\ast_{d\mb{q}}{\cdot \mb{u}}_{c\mb{k}}\right)=\sum^N_{n=1}{u^{\left(n\right)*}_{d\mb{q}}u^{\left(n\right)}_{c\mb{k}}}\] 
corresponds to averaged quantities in Eqs.~\eqref{eq_27_}, \eqref{eq_28_},
\begin{equation} \label{eq_31_} 
 \left({\mb{u}}^\ast_{d\mb{q}}{\cdot \mb{u}}_{c\mb{k}}\right)=\sum^2_{s=1}{\overline{{\psi}^\ast_{d\mb{q}}{\left(\mb{r},s\right)\psi}_{c\mb{k}}\left(\mb{r},s\right)}},                                       
\end{equation} 
or, for ``spinless'' particles,
\begin{equation} \label{eq_32_} 
\left({\mb{u}}^\ast_{d\mb{q}}{\cdot \mb{u}}_{c\mb{k}}\right)=\overline{{\psi}^\ast_{d\mb{q}}{\left(\mb{r}\right)\psi}_{c\mb{k}}\left(\mb{r}\right)}.                                                      
\end{equation} 

For a model with the ``averaged'' Hamiltonian Eq.~\eqref{eq_29_}, the influence of perturbing fields given by the electrodynamic potentials $ \varphi \left(\mb{r},t\right) $ and $ \mb{A}\left(\mb{r},t\right) $ is taken into account by transforming the unperturbed Hamiltonian $ {\hat{H}}_0\left(\hat{\mb{p}}\right)\Rightarrow \hat{H}\left(\hat{\mb{p}},\mb{A},\varphi \right) $ using  Eq.~\eqref{eq_7_}. The equation for the density matrix Eq.~\eqref{eq_13_} with the ``averaged'' Hamiltonian $ \hat{H}\left(\hat{\mb{p}},\mb{A},\varphi \right) $ satisfies the continuity equation \eqref{eq_14_}, in which  
\begin{equation} \label{eq_33_} 
n\left(\mb{r}\right)=g\sum_{c\mb{k}d\mb{q}}{{\mathrm{e}}^{i\left(\mb{k}-\mb{q}\right)\mb{r}}\left({\mb{u}}^\ast_{d\mb{q}}{\cdot \mb{u}}_{c\mb{k}}\right){\rho}_{c\mb{k}d\mb{q}}},                                                                 
\end{equation} 
\begin{equation} \label{eq_34_} 
\mb{j}\left(\mb{r}\right)=-\frac{e}{2}g\sum_{c\mb{k}d\mb{q}}{\left[{\mb{u}}^\ast_{d\mb{q}}{\mathrm{e}}^{-i\mb{qr}}\cdot \left(\hat{\mb{v}}\cdot {\mb{u}}_{c\mb{k}}{\mathrm{e}}^{i\mb{kr}}\right)+\left({\hat{\mb{v}}}^{\mb{*}}\cdot {\mb{u}}^\ast_{d\mb{q}}{\mathrm{e}}^{-i\mb{qr}}\right){\cdot \mb{u}}_{c\mb{k}}{\mathrm{e}}^{i\mb{kr}}\right]}{\rho}_{c\mb{k}d\mb{q}},     
\end{equation} 
where the velocity operator $ \hat{\mb{v}}=\frac{i}{\hbar}\left[\hat{H},\mb{r}\right] $ is the $ N\times N $ matrix, $ g $ is a degeneracy factor which can take into account both spin degeneracy and the presence of identical extreme points in different valleys of the Brillouin zone (see, for example, \cite{katsnelson2012graphene}).

For massless Dirac fermions in Eq.~\eqref{eq_29_} we have $ {\left(\frac{{\partial}^2\hat{E}}{\partial k_i\partial k_j}\right)}_{\mb{k}={\mb{k}}_0}=0 $; therefore, the velocity operator matrix is composed of constant elements and does not involve differentiation:
\[\hat{\mb{v}}=\frac{i}{\hbar}\left[\hat{H},\mb{r}\right]={\frac{i}{{\hbar}^2}\left(\frac{\partial}{\partial \mb{k}}\hat{E}\right)}_{\mb{k}={\mb{k}}_0}\cdot \left[\hat{\mb{p}},\mb{r}\right]={\frac{1}{\hbar}\left(\frac{\partial}{\partial \mb{k}}\hat{E}\right)}_{\mb{k}={\mb{k}}_0}.\] 
Thus, for fermions in Weyl semimetals, graphene and low-energy surface states in topological insulators of the type Bi$_2$Se$_3$, such an ``algebraic'' speed operator is formed by Pauli matrices and is a $ 2\times 2 $ matrix \cite{vafek2014}; for Kane fermions in Cd$_x$Hg$_{1-x}$Te, it is a $ 6\times 6 $ matrix \cite{orlita2014}. In these and similar cases, the products in the expression for the current density Eq.~\eqref{eq_34_} are all algebraic, which leads to a certain simplification (see, for example, \cite{wang2016}). Since in this case we have $ {\mb{u}}^\ast_{d\mb{q}}\left(\hat{\mb{v}}{\mb{u}}_{c\mb{k}}\right)=\left({\hat{\mb{v}}}^{\mb{*}}{\mb{u}}^\ast_{d\mb{q}}\right){\mb{u}}_{c\mb{k}} $, it follows from Eq.~\eqref{eq_34_} that 
\[\mb{j}\left(\mb{r}\right)=-eg\sum_{c\mb{k}d\mb{q}}{{\mathrm{e}}^{i\left(\mb{k}-\mb{q}\right)\mb{r}}\left({\mb{u}}^\ast_{d\mb{q}}\cdot \hat{\mb{v}}\cdot {\mb{u}}_{c\mb{k}}\right){\rho}_{c\mb{k}d\mb{q}}}.\] 

The continuity equation for an ensemble of electrons described by a truncated Hamiltonian is satisfied when 
\[\sum_{cd\mb{k}\mb{q}}{\left({\mb{u}}^\ast_{d\mb{q}}\cdot {\mb{u}}_{c\mb{k}}\right){\mathrm{e}}^{i\left(\mb{k}-\mb{q}\right)\mb{r}}R_{c\mb{k}d\mb{q}}}=0.\] 
After the  derivation similar to Eq.~\eqref{eq_24_}-\eqref{eq_27_} (or using pairs of equations \eqref{eq_27_},\eqref{eq_31_} or \eqref{eq_28_},\eqref{eq_32_}) we obtain the following expression for the relaxation matrix written in the form of Eq.~\eqref{eq_25_}: 
\begin{equation} \label{eq_35_} 
G_{c\mb{k}d\mb{q}}=\frac{1}{G_{d\,-\mb{q}\,c\,-\mb{k}}}=\frac{\left({\mb{u}}^\ast_{c-\mb{k}}{\cdot \mb{u}}_{d-\mb{q}}\right)}{\left({\mb{u}}^\ast_{d\mb{q}}{\cdot \mb{u}}_{c\mb{k}}\right)}.                                                  
\end{equation} 
Note that for a wide class of systems the following condition is satisfied: 
\begin{equation} \label{eq_36_} 
{\left|G_{c\mb{k}d\mb{q}}\right|}^2=1. 
\end{equation} 
This is similar to  a non-averaged system described by Bloch eigenfunctions, when  the relaxation operator is given by Eq.~\eqref{eq_22_}, and the factor $ {\mathrm{e}}^{i\left({\varphi}_{d\mb{q}}-{\varphi}_{c\mb{k}}\right)} $ replaces the coefficient $ G_{c\mb{k}d\mb{q}} $. In Appendix A we will show that to satisfy the condition \eqref{eq_36_} it is sufficient (although not necessary) to have a TRS effective Hamiltonian $ {\hat{H}}_0\left(\hat{\mb{p}}\right) $.  Appendix B contains an example of a system (a Weyl semimetal) with broken time-reversal symmetry, for which the condition \eqref{eq_36_} is nevertheless satisfied. Appendix C shows that when calculating the linear susceptibility in the limit of a uniform external field, violating the condition \eqref{eq_36_} does not affect the result.

Thus, Eqs.~\eqref{eq_25_}, \eqref{eq_27_}, \eqref{eq_35_} 
define a relatively simple relaxation operator, the use of which in the master equations preserves the  continuity equation for the observables.
Note that the relaxation operator Eq.~\eqref{eq_25_} cannot be reduced to the standard form Eq.~\eqref{eq_2_} by a formal replacement $ G_{c\mb{k}d\mb{q}}=0 $. Accordingly, the response of the medium obtained using the relaxation operator Eq.~\eqref{eq_25_} does not reduce to the one obtained on the basis of the standard relaxation model by replacing $ G_{c\mb{k}d\mb{q}}=0 $. This is because the coefficients $ G_{c\mb{k}d\mb{q}} $ determined by the properties of the eigenstates of the Hamiltonian obey Eq.~\eqref{eq_27_} or Eq.~\eqref{eq_35_}. Therefore, formally setting $G_{c\mb{k}d\mb{q}}=0$ for any transition, we automatically have $ G_{d\,-\mb{q}\,c\,-\mb{k}}=\infty $.

\section{Generalization of the Lindhard formula in a dissipative system in two dimensions}
\label{sec_lindhard}

\subsection{Screening effect in a monolayer}

Let the monolayer with charge carriers be located in the plane $z=0$. In the region $ z<0 $, there is a substrate with a dielectric constant $ \varepsilon $. Consider the electric field potential $ \mathrm{\Phi}\left(\mb{r},z,t\right) $, where the vector $ \mb{r} $ belongs to the \textit{xy} plane. We write the potential as an  expansion in 2D Fourier harmonics, 
\[\mathrm{\Phi} = \int{d\omega}\mathop{\int\!\!\!\!\int}{{\mathrm{\Phi}}_{\mb{\kappa}\omega}\left(z\right){\mathrm{e}}^{i\mb{\kappa}\mb{r}-i\omega t}d^2\kappa}.\] 
The complex amplitudes of the field harmonics in the layer plane are given by  $ {\mb{E}}_{\mb{\kappa}\omega}=-i\mb{\kappa}\mathrm{\Phi}_{\mb{\kappa}\omega}\left(0\right) $. Hereafter, to simplify the expressions, we will use the notation $ {\mathrm{\Phi}}_{\mb{\kappa}\omega} $ instead of $ {\mathrm{\Phi}}_{\mb{\kappa}\omega}\left(0\right) $. 

Let $ \chi \left(\omega ,\mb{\kappa}\right) $ be the 2D linear susceptibility of a layer, which determines its surface polarization excited by a field harmonic: 
\[{\mb{P}}_{\mb{\kappa}\omega}=-i\mb{\kappa}\chi \left(\omega ,\mb{\kappa}\right){\mathrm{\Phi}}_{\mb{\kappa}\omega}.\] 
Harmonics of the surface charge are related to the harmonics of surface polarization by $ Q_{\mb{\kappa}\omega}=-i\mb{\kappa}{\mb{P}}_{\mb{\kappa}\omega} $, from which 
\begin{equation} \label{eq_37_} 
Q_{\mb{\kappa}\omega}=-{\kappa}^2\chi \left(\omega ,\mb{\kappa}\right){\mathrm{\Phi}}_{\mb{\kappa}\omega}.                                                         
\end{equation} 

We will seek a response to the external potential $ \mathrm{\Phi} $, taking into account the excitation of the self-consistent potential $ \delta \mathrm{\Phi} $:
\begin{equation} \label{eq_38_} 
Q_{\mb{\kappa}\omega}=-{\kappa}^2\chi \left(\omega ,\mb{\kappa}\right)\left({\mathrm{\Phi}}_{\mb{\kappa}\omega}+\delta {\mathrm{\Phi}}_{\mb{\kappa}\omega}\right).                                              
\end{equation} 
The Poisson equation outside the monolayer gives 
\[\left(-{\kappa}^2+\frac{{\partial}^2}{\partial z^2}\right){\delta \mathrm{\Phi}}_{\mb{\kappa}\omega}\left(z\right)=0.\] 
Its continuous solution along the z axis is 
\[{\delta \mathrm{\Phi}}_{\mb{\kappa}\omega}\left(z\right) = \delta {\mathrm{\Phi}}_{\mb{\kappa}\omega}{\mathrm{e}}^{\mp \kappa z},\] 
where the upper and lower sign corespond to the upper and lower half-spaces. The value of $ \delta {\mathrm{\Phi}}_{\mb{\kappa}\omega} $ can be determined using the Gauss theorem: 
\begin{equation} \label{eq_39_} 
\kappa \delta {\mathrm{\Phi}}_{\mb{\kappa}\omega}+\varepsilon \kappa \delta {\mathrm{\Phi}}_{\mb{\kappa}\omega}=4\pi Q_{\mb{\kappa}\omega}.                                                  
\end{equation} 
As a result, we get from Eqs.~\eqref{eq_38_}, \eqref{eq_39_} 
\begin{equation} \label{eq_40_} 
{\mathrm{\Phi}}^{\left(\mathrm{scr}\right)}_{\mb{\kappa}\omega} = \frac{{\mathrm{\Phi}}_{\mb{\kappa}\omega}}{1+\frac{\kappa}{1+\varepsilon}4\pi \chi \left(\omega ,\mb{\kappa}\right)},                                                                 
\end{equation} 
where $ {\mathrm{\Phi}}^{\left(\mathrm{scr}\right)}_{\mb{\kappa}\omega}={\mathrm{\Phi}}_{\mb{\kappa}\omega}+\delta {\mathrm{\Phi}}_{\mb{\kappa}\omega} $ is the harmonic of a screened potential. Note that equating the expression in the denominator Eq.~\eqref{eq_40_} to zero gives the dispersion equation for a 2D plasmon supported by the monolayer \cite{yao2014}: $ 1+\frac{\kappa}{1+\varepsilon}4\pi \chi \left(\omega ,\mb{\kappa}\right)=0 $.

Let us show that Eq.~\eqref{eq_40_} corresponds exactly to the Lindhard formula for a 2D system in the absence of dissipation \cite{haug2009quantum}:
\begin{equation} \label{eq_41_} 
{\mathrm{\Phi}}^{\left(\mathrm{scr}\right)}_{\mb{\kappa}\omega} = \frac{{\mathrm{\Phi}}_{\mb{\kappa}\omega}}{1- \frac{2}{1+\varepsilon}{\mathrm{\Phi}}_{0\mb{\kappa}}g\sum_{\alpha \beta}{\frac{\left(f_{\alpha}-f_{\beta}\right){\left|{\left({\mathrm{e}}^{i\mb{\kappa}\mb{r}}\right)}_{\alpha \beta}\right|}^2}{E_{\alpha}-E_{\beta}-\hbar \omega}}},                                                        
\end{equation} 
where $ {\mathrm{\Phi}}_{0\mb{\kappa}} $ is a spatial 2D harmonic of the interaction potential of point charges $ {e^2}/{r} $, 
\[{\mathrm{\Phi}}_{0\mb{\kappa}}=\frac{1}{4{\pi}^2}\int_{\infty}{\frac{e^2}{r}}{\mathrm{e}}^{-i\mb{\kappa}\mb{r}}d^2r=\frac{e^2}{2\pi \kappa},\] 
 $ f_{\alpha} $ and $ E_{\alpha} $ are the population and energy of quasiparticles corresponding to the state $ |\alpha \rangle $, and $ g $ is the degeneracy factor.

To compare Eq.~\eqref{eq_40_} and Eq.~\eqref{eq_41_}, we need to obtain an expression for the susceptibility $ \chi \left(\omega ,\mb{\kappa}\right) $. We use the equation for the complex amplitude of the linear perturbation of the density matrix $ {{\rho}_{\alpha \beta}=\tilde{\rho}}_{\alpha \beta}{\mathrm{e}}^{-i\omega t} $ under the action of the harmonic of the potential $ {\mathrm{\Phi}}_{\mb{\kappa}\omega}{\mathrm{e}}^{i\mb{\kappa}\mb{r}-i\omega t} $,  
\begin{equation} \label{eq_42_} 
-i\omega {\tilde{\rho}}_{\alpha \beta}+i\frac{E_{\alpha}-E_{\beta}}{\hbar}{\tilde{\rho}}_{\alpha \beta}=-\frac{i}{\hbar}e{\mathrm{\Phi}}_{\mb{\kappa}\omega}{\left({\mathrm{e}}^{i\mb{\kappa}\mb{r}}\right)}_{\alpha \beta}\left(f_{\alpha}-f_{\beta}\right),                               
\end{equation} 
which yields 
\begin{equation} \label{eq_43_} 
{\tilde{\rho}}_{\alpha \beta}=-e\frac{{\mathrm{\Phi}}_{\mb{\kappa}\omega}{\left({\mathrm{e}}^{i\mb{\kappa}\mb{r}}\right)}_{\alpha \beta}\left(f_{\alpha}-f_{\beta}\right)}{E_{\alpha}-E_{\beta}-\hbar \omega}.                                                      
\end{equation} 
It follows from the first of Eqs.~\eqref{eq_12_} that 
\begin{equation} \label{eq_44_} 
Q_{\mb{\kappa}\omega} =- \frac{eg}{4{\pi}^2}\sum_{\alpha \beta}{{\left({\mathrm{e}}^{-i\mb{\kappa}\mb{r}}\right)}_{\beta \alpha}{\tilde{\rho}}_{\alpha \beta}}.                                                  
\end{equation} 
Substituting Eqs.~\eqref{eq_43_}, \eqref{eq_44_} into Eq.~\eqref{eq_37_} and using the relation $ {\left({\mathrm{e}}^{-i\mb{\kappa}\mb{r}}\right)}_{\beta \alpha}{\left({\mathrm{e}}^{i\mb{\kappa}\mb{r}}\right)}_{\alpha \beta}={\left|{\left({\mathrm{e}}^{i\mb{\kappa}\mb{r}}\right)}_{\alpha \beta}\right|}^2 $ we obtain 
\begin{equation} \label{eq_45_} 
\chi \left(\omega ,\mb{\kappa}\right)=-\frac{e^2g}{4{\pi}^2{\kappa}^2}\sum_{\alpha \beta}{\frac{\left(f_{\alpha}-f_{\beta}\right){\left|{\left({\mathrm{e}}^{i\mb{\kappa}\mb{r}}\right)}_{\alpha \beta}\right|}^2}{E_{\alpha}-E_{\beta}-\hbar \omega}}.                                             
\end{equation} 
It is easy to see that the substitution of Eq.~\eqref{eq_45_} into Eq.~\eqref{eq_40_} leads to Eq.~\eqref{eq_41_}.

Another way to derive the linear susceptibility is to calculate the Fourier harmonic of the current $ {\mb{j}}_{\mb{\kappa}\omega} $, which follows from the second of  Eqs.~\eqref{eq_12_}:
\begin{equation} \label{eq_46_} 
{\mb{j}}_{\mb{\kappa}\omega} =- \frac{eg}{4{\pi}^2}\sum_{\alpha \beta}{{\frac{1}{2}\left({\mathrm{e}}^{-i\mb{\kappa}\mb{r}}\hat{\mb{v}}+\hat{\mb{v}}{\mathrm{e}}^{-i\mb{\kappa}\mb{r}}\right)}_{\beta \alpha}{\tilde{\rho}}_{\alpha \beta}}.                                           
\end{equation} 
Using the identity $ {\left(\frac{\mathrm{\nabla}u\cdot \hat{\mb{v}}+\hat{\mb{v}}\cdot \mathrm{\nabla}u}{2}\right)}_{\beta \alpha}=i\frac{E_{\beta}-E_{\alpha}}{\hbar}u_{\beta \alpha} $ which holds for any function $ u\left(\mb{r}\right) $ (see, for example, \cite{wang2016}), from Eqs.~\eqref{eq_46_} and \eqref{eq_43_} we obtain the expression for the conductivity 
\begin{equation} \label{eq_47_} 
\sigma \left(\omega ,\mb{\kappa}\right)=i\frac{e^2g}{4{\pi}^2{\kappa}^2}\sum_{\alpha \beta}{\frac{E_{\alpha}-E_{\beta}}{\hbar}\times \frac{\left(f_{\alpha}-f_{\beta}\right){\left|{\left({\mathrm{e}}^{i\mb{\kappa}\mb{r}}\right)}_{\alpha \beta}\right|}^2}{E_{\alpha}-E_{\beta}-\hbar \omega}}.                                             
\end{equation} 
Using the relation $ \sum_{\alpha}{{\left({\mathrm{e}}^{\mp i\mb{\kappa}\mb{r}}\right)}_{\beta \alpha}{\left({\mathrm{e}}^{\pm i\mb{\kappa}\mb{r}}\right)}_{\alpha \beta}}=1 $, which is valid for any index $ \beta $, Eq.~\eqref{eq_47_} can be reduced to the following form, 
\[\sigma \left(\omega ,\mb{\kappa}\right)=i\omega \frac{e^2g}{4{\pi}^2{\kappa}^2}\sum_{\alpha \beta}{\frac{\left(f_{\alpha}-f_{\beta}\right){\left|{\left({\mathrm{e}}^{i\mb{\kappa}\mb{r}}\right)}_{\alpha \beta}\right|}^2}{E_{\alpha}-E_{\beta}-\hbar \omega}}=-i\omega \chi \left(\omega ,\mb{\kappa}\right),\] 
which corresponds to the fundamental relationship Eq.~\eqref{eq_3_}.

\subsection{Accounting for relaxation within the standard model}

 Using the standard relaxation operator defined by Eq.~\eqref{eq_2_} in the equation for the perturbation of the density matrix Eq.~\eqref{eq_42_} results in the substitution $ \omega \to \omega +i{\gamma}_{\alpha \beta} $ in the corresponding relations Eqs.~\eqref{eq_45_}, \eqref{eq_47_}:
\begin{equation} \label{eq_48_} 
\chi \left(\omega ,\mb{\kappa}\right)=-\frac{e^2}{4{\pi}^2{\kappa}^2}g\sum_{\alpha \beta}{\frac{\left(f_{\alpha}-f_{\beta}\right){\left|{\left({\mathrm{e}}^{i\mb{\kappa}\mb{r}}\right)}_{\alpha \beta}\right|}^2}{E_{\alpha}-E_{\beta}-\hbar \omega -i\hbar {\gamma}_{\alpha \beta}}},                                                        
\end{equation} 
\begin{equation} \label{eq_49_} 
\sigma \left(\omega ,\mb{\kappa}\right)=\frac{e^2g}{4{\pi}^2{\kappa}^2}\sum_{\alpha \beta}{\left(i\omega -{\gamma}_{\alpha \beta}\right)\frac{\left(f_{\alpha}-f_{\beta}\right){\left|{\left({\mathrm{e}}^{i\mb{\kappa}\mb{r}}\right)}_{\alpha \beta}\right|}^2}{E_{\alpha}-E_{\beta}-\hbar \omega -i\hbar {\gamma}_{\alpha \beta}}} 
\end{equation} 
Obviously, this solution obtained using the relaxation operator in the form of Eq.~\eqref{eq_2_} violates Eq.~\eqref{eq_3_}. This can lead to significant errors in the low-frequency range (see, for example, \cite{tokman2013,zhang2014,tokman2009}).

\subsection{Accounting for relaxation with the modified relaxation operator}

Let us consider a system with quasi-particle states $ |\alpha \rangle =|c,\mb{k}\rangle$ using the relaxation operator Eq.~\eqref{eq_25_}, which preserves the average number of particles. The density matrix equations become 
\begin{equation} \label{eq_50_}
-i\omega {\tilde{\rho}}_{c\mb{k}d\mb{q}}+i\frac{E_{c\mb{k}}-E_{d\mb{q}}}{\hbar}{\tilde{\rho}}_{c\mb{k}d\mb{q}}=-\frac{i}{\hbar}e{\mathrm{\Phi}}_{\omega ;c\mb{k}d\mb{q}}\left(f_{c\mb{k}}-f_{d\mb{q}}\right)-{\gamma}_{c\mb{k}d\mb{q}}\left({\tilde{\rho}}_{c\mb{k}d\mb{q}}-{G_{c\mb{k}d\mb{q}}\tilde{\rho}}_{d\,-\mb{q}\,c\,-\mb{k}}\right),
\end{equation}
\begin{equation} \label{eq_51_}
-i\omega {\tilde{\rho}}_{d\,-\mb{q}\,c\,-\mb{k}}+i\frac{E_{d\mb{q}}-E_{c\mb{k}}}{\hbar}{\tilde{\rho}}_{d\,-\mb{q}\,c\,-\mb{k}}=-\frac{i}{\hbar}e{\mathrm{\Phi}}_{\omega ;d-\mb{q}\,c-\mb{k}}\left(f_{d-\mb{q}}-f_{c-\mb{k}}\right)-{\gamma}_{cd\mb{k}\mb{q}}\left({\tilde{\rho}}_{d\,-\mb{q}\,c\,-\mb{k}}-G_{d\,-\mb{q}\,c\,-\mb{k}}{\tilde{\rho}}_{c\mb{k}d\mb{q}}\right),
\end{equation}
where $ {\mathrm{\Phi}}_{\omega ;c\mb{k}d\mb{q}}={\mathrm{\Phi}}_{\mb{\kappa}\omega}{\left({\mathrm{e}}^{i\mb{\kappa}\mb{r}}\right)}_{c\mb{k}d\mb{q}} $, $ {\gamma}_{c\mb{k}d\mb{q}}={\gamma}_{d\,-\mb{q}\,c\,-\mb{k}} $, and energy $ E_{c\mb{k}} $ does not depend on the sign of $\mb{k}$. From Eqs.~\eqref{eq_50_}, \eqref{eq_51_}, taking into account the relationship $ G_{c\mb{k}d\mb{q}}G_{d\,-\mb{q}\,c\,-\mb{k}}=1 $ (see Eq.~\eqref{eq_35_}), we obtain  
\begin{align}
{\tilde{\rho}}_{c\mb{k}d\mb{q}} & = -e\frac{\left(E_{d\mb{q}}-E_{c\mb{k}}-\hbar \omega \right){\mathrm{\Phi}}_{\omega ;c\mb{k}d\mb{q}}\left(f_{c\mb{k}}-f_{d\mb{q}}\right)}{{\hbar}^2{\omega}^2+2{i\hbar}^2\omega {\gamma}_{c\mb{k}d\mb{q}}-{\left(E_{c\mb{k}}-E_{d\mb{q}}\right)}^2}  \nonumber \\
&  + i\hbar {\gamma}_{c\mb{k}d\mb{q}}e\frac{{\mathrm{\Phi}}_{\omega ;c\mb{k}d\mb{q}}\left(f_{c\mb{k}}-f_{d\mb{q}}\right)-G_{c\mb{k}d\mb{q}}{\mathrm{\Phi}}_{\omega ;d-\mb{q}\,c-\mb{k}}\left(f_{c-\mb{k}}-f_{d-\mb{q}}\right)}{{\hbar}^2{\omega}^2+2{i\hbar}^2\omega {\gamma}_{c\mb{k}d\mb{q}}-{\left(E_{c\mb{k}}-E_{d\mb{q}}\right)}^2}.  
  \label{eq_52_} 
\end{align} 

Within the averaged description, the following relationship is satisfied,  
\begin{equation} \label{eq_53_} 
{\mathrm{\Phi}}_{\omega ;c\mb{k}d\mb{q}}={\mathrm{\Phi}}_{\mb{\kappa}\omega}{\delta}_{\mb{k}\left(\mb{q}+\mb{\kappa}\right)}\left({\mb{u}}^\ast_{c\mb{k}}\cdot {\mb{u}}_{d\mb{q}}\right),                                                   
\end{equation} 
which gives, after taking Eqs.~\eqref{eq_35_} into account,  
\begin{equation} \label{eq_54_} 
{\mathrm{\Phi}}_{\omega ;d-\mb{q}\,c-\mb{k}}={\mathrm{\Phi}}_{\omega ;c\mb{k}d\mb{q}}G^\ast_{c\mb{k}d\mb{q}} .                                                     
\end{equation} 
Using Eqs.~\eqref{eq_53_}, \eqref{eq_54_} and assuming that the populations are determined only by the energies of states, we transform the second term on the right-hand side of Eq.~\eqref{eq_52_} into  
\[i\hbar {\gamma}_{c\mb{k}d\mb{q}}e{\mathrm{\Phi}}_{\mb{\kappa}\omega}{\delta}_{\mb{k}\left(\mb{q}+\mb{\kappa}\right)}\left({\mb{u}}^\ast_{c\mb{k}}\cdot {\mb{u}}_{d\mb{q}}\right)\frac{\mathrm{1}-{\left|G_{c\mb{k}d\mb{q}}\right|}^2}{{\hbar}^2{\omega}^2+2{i\hbar}^2\omega {\gamma}_{c\mb{k}d\mb{q}}-{\left(E_{c\mb{k}}-E_{d\mb{q}}\right)}^2}\left(f_{c\mb{k}}-f_{d\mb{q}}\right).\] 
Under the condition \eqref{eq_36_} the second term on the right-hand side of Eq.~\eqref{eq_52_} is equal to zero. As a result, we obtain the following expression for the perturbation of the density matrix, 
\begin{equation} \label{eq_55_} 
{\tilde{\rho}}_{c\mb{k}d\mb{q}}=-e\frac{\left(E_{c\mb{k}}-E_{d\mb{q}}+\hbar \omega \right){\mathrm{\Phi}}_{\omega ;c\mb{k}d\mb{q}}\left(f_{c\mb{k}}-f_{d\mb{q}}\right)}{{\left(E_{c\mb{k}}-E_{d\mb{q}}\right)}^2-{\hbar}^2{\omega}^2-2{i\hbar}^2\omega {\gamma}_{c\mb{k}d\mb{q}}}  .
\end{equation}

As we noted at the end of the section II, for a wide class of systems the coefficients $\left|G_{c\mb{k}d\mb{q}}\right|=1$  (condition \eqref{eq_36_} is satisfied). We assume this to be true here. Moreover, it will be shown in Appendix C that the derivation of the linear susceptibility in the limit of a uniform external field for an arbitrary value of $\left|G_{c\mb{k}d\mb{q}}\right|$ gives the same result as for $\left|G_{c\mb{k}d\mb{q}}\right|=1$.

%Note that the relaxation operator Eq.~\eqref{eq_25_} cannot be reduced to the standard form Eq.~\eqref{eq_2_} by a formal replacement $ G_{c\mb{k}d\mb{q}}=0 $. Accordingly, the response of the medium obtained using the relaxation operator Eq.~\eqref{eq_25_} does not reduce to the one obtained on the basis of the standard relaxation model by replacing $ G_{c\mb{k}d\mb{q}}=0 $. This is because the coefficients $ G_{c\mb{k}d\mb{q}} $ determined by the properties of the eigenstates of the Hamiltonian obey Eq.~\eqref{eq_27_} or Eq.~\eqref{eq_28_}. Therefore, formally setting $G_{c\mb{k}d\mb{q}}=0$ for any transition, we automatically have $ G_{d\,-\mb{q}\,c\,-\mb{k}}=\infty $. 

The expression for the amplitude of monochromatic oscillations of the charge is calculated after substituting Eq.~\eqref{eq_55_} into  Eq.~\eqref{eq_44_}. As a result, taking into account Eq.~\eqref{eq_37_}, we obtain the expression for the susceptibility,
\begin{equation} \label{eq_56_} 
\chi \left(\omega ,\mb{\kappa}\right)=-\frac{e^2g}{4{\pi}^2{\kappa}^2}\sum_{cd\mb{k}\mb{q}}\frac{\left(f_{c\mb{k}}-f_{d\mb{q}}\right){{\delta}_{\mb{k}\left(\mb{q}+\mb{\kappa}\right)}\left|{\mb{u}}^\ast_{c\mb{k}}\cdot {\mb{u}}_{d\mb{q}}\right|}^2\left(E_{c\mb{k}}-E_{d\mb{q}}+\hbar \omega \right)}{{{\left(E_{c\mb{k}}-E_{d\mb{q}}\right)}^2-\hbar^2\omega^2-2{i\hbar}^2\omega {\gamma}_{c\mb{k}d\mb{q}}}}.                           
\end{equation} 
It is easy to verify that as $ {\gamma}_{c\mb{k}d\mb{q}}\to 0 $, the expression Eq.~\eqref{eq_56_} transforms into the dissipationless formula Eq.~\eqref{eq_45_}.

The resulting expression for the susceptibility can be simplified, given the identity 
\begin{equation} \label{eq_57_} 
\sum_{cd\mb{k}\mb{q}}\frac{\left(f_{c\mb{k}}-f_{d\mb{q}}\right){{\delta}_{\mb{k}\left(\mb{q}+\mb{\kappa}\right)}\left|{\mb{u}}^\ast_{c\mb{k}}\cdot {\mb{u}}_{d\mb{q}}\right|}^2}{{{\left(E_{c\mb{k}}-E_{d\mb{q}}\right)}^2-\hbar^2{\omega}^2-2{i\hbar}^2\omega {\gamma}_{c\mb{k}d\mb{q}}}}=0.                                          
\end{equation} 
To prove Eq.~\eqref{eq_57_} we make a replacement $ c\mb{k} \leftrightarrow d\mb{q} $ in the sum and obtain 
\[\sum_{cd\mb{k}\mb{q}}\frac{\left(f_{c\mb{k}}-f_{d\mb{q}}\right){{\delta}_{\mb{k}\left(\mb{q}+\mb{\kappa}\right)}\left|{\mb{u}}^\ast_{c\mb{k}}\cdot {\mb{u}}_{d\mb{q}}\right|}^2}{{{\left(E_{c\mb{k}}-E_{d\mb{q}}\right)}^2-\hbar^2{\omega}^2-2{i\hbar}^2\omega {\gamma}_{c\mb{k}d\mb{q}}}}=\sum_{cd\mb{k}\mb{q}}\frac{\left(f_{c\mb{k}}-f_{d\mb{q}}\right){\left|{\mb{u}}^\ast_{c\mb{k}}\cdot {\mb{u}}_{d\mb{q}}\right|}^2\frac{{\delta}_{\mb{k}\left(\mb{q}+\mb{\kappa}\right)}-{\delta}_{\mb{k}\left(\mb{q}-\mb{\kappa}\right)}}{2}}{{{\left(E_{c\mb{k}}-E_{d\mb{q}}\right)}^2-\hbar^2{\omega}^2-2{i\hbar}^2\omega {\gamma}_{c\mb{k}d\mb{q}}}}.\] 
Under the condition $ \left|G_{c\mb{k}d\mb{q}}\right|=1 $ we always have  $ {\left|{\mb{u}}^\ast_{c\mb{k}}\cdot {\mb{u}}_{d\mb{q}}\right|}^2={\left|{\mb{u}}^\ast_{c-\mb{k}}\cdot {\mb{u}}_{d-\mb{q}}\right|}^2 $. In this case, the right-hand side of the last expression can be represented as the difference of identical sums, so that Eq.~\eqref{eq_57_} is satisfied. As result we get
\begin{equation} \label{eq_58_} 
\chi \left(\omega ,\mb{\kappa}\right)=-\frac{e^2g}{4{\pi}^2{\kappa}^2}\sum_{cd\mb{k}\mb{q}}\frac{\left(E_{c\mb{k}}-E_{d\mb{q}}\right)\left(f_{c\mb{k}}-f_{d\mb{q}}\right){{\delta}_{\mb{k}\left(\mb{q}+\mb{\kappa}\right)}\left|{\mb{u}}^\ast_{c\mb{k}}\cdot {\mb{u}}_{d\mb{q}}\right|}^2}{{{\left(E_{c\mb{k}}-E_{d\mb{q}}\right)}^2-\hbar^2{\omega}^2-2{i\hbar}^2\omega {\gamma}_{c\mb{k}d\mb{q}}}}.                           \end{equation} 
For an independent derivation of the conductivity, we use  Eqs.~\eqref{eq_46_}, \eqref{eq_55_} to obtain
\[\sigma \left(\omega ,\mb{k}\right)=i\frac{e^2g}{4{\pi}^2{\kappa}^2}\sum_{c\mb{k}d\mb{q}}{\frac{E_{c\mb{k}}-E_{d\mb{q}}}{\hbar}\times \frac{\left(E_{c\mb{k}}-E_{d\mb{q}}+\hbar \omega \right){{\delta}_{\mb{k}\left(\mb{q}+\mb{\kappa}\right)}\left|{\mb{u}}^\ast_{c\mb{k}}\cdot {\mb{u}}_{d\mb{q}}\right|}^2\left(f_{c\mb{k}}-f_{d\mb{q}}\right)}{{\left(E_{c\mb{k}}-E_{d\mb{q}}\right)}^2-{\hbar}^2{\omega}^2-2{i\hbar}^2\omega {\gamma}_{c\mb{k}d\mb{q}}}}.\] 
Then we use the relation 
\[\sum_{cd\mb{k}\mb{q}}{\frac{{\left(E_{c\mb{k}}-E_{d\mb{q}}\right)}^2\left(f_{c\mb{k}}-f_{d\mb{q}}\right){{\delta}_{\mb{k}\left(\mb{q}+\mb{\kappa}\right)}\left|{\mb{u}}^\ast_{c\mb{k}}\cdot {\mb{u}}_{d\mb{q}}\right|}^2}{{{\left(E_{c\mb{k}}-E_{d\mb{q}}\right)}^2-\hbar}^2{\omega}^2-2{i\hbar}^2\omega {\gamma}_{c\mb{k}d\mb{q}}}}=0,\] 
which proof is completely analogous to that of Eq.~\eqref{eq_57_}. As a result, we arrive at 
\[\sigma \left(\omega ,\mb{k}\right)=i\omega \frac{e^2g}{4{\pi}^2{\kappa}^2}\sum_{c\mb{k}d\mb{q}}{\frac{\left(E_{c\mb{k}}-E_{d\mb{q}}\right)\left(f_{c\mb{k}}-f_{d\mb{q}}\right){{\delta}_{\mb{k}\left(\mb{q}+\mb{\kappa}\right)}\left|{\mb{u}}^\ast_{c\mb{k}}\cdot {\mb{u}}_{d\mb{q}}\right|}^2}{{\left(E_{c\mb{k}}-E_{d\mb{q}}\right)}^2-{\hbar}^2{\omega}^2-2{i\hbar}^2\omega {\gamma}_{c\mb{k}d\mb{q}}}}=-i\omega \chi \left(\omega ,\mb{\kappa}\right).\] 
As we see, in this case the relationship \eqref{eq_3_} is always satisfied; therefore, it suffices to analyze the properties of the expression \eqref{eq_58_}, which determines the value of $ \chi(\omega,\mb{\kappa}) $.

An important feature of the expression \eqref{eq_58_} is that the transition to the limit of a constant field is nontrivial. Depending on the order in which we take the limits, the relations obtained in the limit $ \omega \to 0 $ allow us to describe both the response of an equilibrium quasi-closed system to a perturbing potential and the ohmic conductivity of an open system.

 For a nonzero value of $ \mb{\kappa} $ in the limit $ \omega =0 $, Eq.~\eqref{eq_58_} defines the real value of $ \chi $, independent on the relaxation constants. If the quantities $ f_{c\mb{k}} $ correspond to the equilibrium distribution in the absence of external fields, then expression \eqref{eq_58_} corresponds to the equilibrium state of the system placed in the potential field, treating the electron-field interaction as a perturbation.  This result corresponds to the correct limit of a stationary response to an external non-uniform constant perturbing field, since such a response itself should not depend on the mechanism and rate of relaxation.

We will arrive at a different result if we first take the  limit $ \mb{\kappa}\to 0 $. In this case we obtain
\begin{equation} \label{eq_59_} 
\chi \left(\omega ,0\right)=-\frac{e^2g}{4{\pi}^2}\sum_{c\neq d,\mb{k}}{\frac{{\left|{\mb{n}\cdot}{\mb{r}}_{c\mb{k}d\mb{k}}\right|}^2\left(f_{c\mb{k}}-f_{d\mb{k}}\right)\left(E_{c\mb{k}}-E_{d\mb{k}}\right)}{{{\left(E_{c\mb{k}}-E_{d\mb{k}}\right)}^2-\hbar}^2{\omega}^2-2{i\hbar}^2\omega {\gamma}_{c\mb{k}d\mb{k}}}}+\frac{e^2g}{4{\pi}^2}\sum_{c\mb{k}}{\frac{\left(\mb{n}\frac{\partial}{\partial \mb{k}}E_{c\mb{k}}\right)\left(\mb{n}\frac{\partial}{\partial \mb{k}}f_{c\mb{k}}\right)  }{{\hbar}^2{\omega}^2+2{i\hbar}^2\omega {\gamma}_{c\mb{k}c\mb{k}}}},            
\end{equation} 
where $ {\mb{r}}_{c\mb{k}d\mb{k}}={\mb{u}}^\ast_{c\mb{k}}i\frac{\partial}{\partial \mb{k}}{\mb{u}}_{d\mb{k}} $ is the matrix element of the coordinate operator defined in the $\mb{k}$-representation for a ``direct'' transition; $ \mb{n}=\frac{\mb{\kappa}}{\left|\mb{\kappa}\right|} $.

It is easy to verify that when we use Eq.~\eqref{eq_59_} to determine the conductivity $ \sigma \left(\omega ,0\right)=-i\omega \chi \left(\omega ,0\right) $, the second term on the right-hand side of Eq.~\eqref{eq_59_} defines the standard intraband Drude conductivity, up  to a factor of 2 in the definition of a relaxation constant. 

To summarize, if the limit $ \omega \to 0 $ is taken after taking the limit $ \mb{\kappa}\to 0 $, then a physically transparent result is obtained: interband transitions determine the susceptibility which does not depend on the relaxation constant, whereas intraband transitions determine the finite Drude conductivity which depends on the relaxation constant. 
However, this conclusion is valid for a finite band gap only. Otherwise (as, for example, in graphene), a more complex contribution of interband transitions is possible.

Note that for $ \omega \to 0 $, a continuous transition from the equilibrium ``current-free'' solution given by Eq.~\eqref{eq_58_} to the ohmic conductivity given by Eq.~\eqref{eq_59_} is possible only for a problem with boundary conditions.

%Appendix C shows that when calculating the linear susceptibility in the limit of a uniform external field for an arbitrary value of $ {\left|G_{c\mb{k}d\mb{q}}\right|}^2 $, the result is the same as for $ {\left|G_{c\mb{k}d\mb{q}}\right|}^2=1 $.

The resonance denominator in the expression for the susceptibility Eq.\eqref{eq_58_} corresponds to a classical harmonic oscillator with friction; this structure of the electrodynamic response is typical for the relaxation operator with an antisymmetric structure like Eq.~\eqref{eq_5_} \cite{tokman2013, zhang2014}.

For the simplest open systems that violate time reversal  symmetry, for example free particles or a harmonic oscillator in a magnetic field,  expressions similar to Eq.\eqref{eq_58_} were obtained in \cite{tokman2013,zhang2014}. The corresponding expressions for the phenomenological relaxation operator were derived there imposing the requirement of the gauge invariance of the electrodynamic response \cite{tokman2013, tokman2009}. In the present case, no additional requirements, except for the conservation of the particle number, were imposed. As noted above, all the necessary information about the system is ``hidden'' in the form of specific Bloch functions $ {\psi}_{c\mb{k}}\left(\mb{r},s\right) $ or, within the simplified description, in the form of vectors $ {\mb{u}}_{c\mb{k}} $. When used in Eqs.~\eqref{eq_25_}, \eqref{eq_27_} or Eqs.~\eqref{eq_25_}, \eqref{eq_35_} respectively, these sets of state functions unambiguously define the correct phenomenological relaxation operator.

%%%%%%%%%%%%%%%%%%%%%%%%%%

\section{Application to graphene}
\label{sec_comparison}

\subsection{General considerations}

In this section, we compare the results for the dielectric response of graphene obtained with the modified relaxation operator Eq.~\eqref{eq_25_} and the standard relaxation operator Eq.~\eqref{eq_2_}. One has to choose for which particular quantity to carry out the comparison: the surface susceptibility $ \chi \left(\omega ,\mb{\kappa}\right) $ or the surface conductivity $ \sigma \left(\omega ,\mb{\kappa}\right) $, since using the standard relaxation operator Eq.~\eqref{eq_2_} violates Eq.~\eqref{eq_3_}. Whenever the standard relaxation operator is used, we introduce the notation $ {\chi}^{\left(st\right)}\left(\omega ,\mb{\kappa}\right) $ and $ {\sigma}^{\left(st\right)}\left(\omega ,\mb{\kappa}\right) $.

For quasiparticle states $ |\alpha \rangle =|c,\mb{k}\rangle $ the expressions Eqs.~\eqref{eq_48_}, \eqref{eq_49_} have the form 
\begin{equation} \label{eq_60_} 
{\chi}^{\left(st\right)}\left(\omega ,\mb{\kappa}\right)=-\frac{e^2g}{4{\pi}^2{\kappa}^2}\sum_{cd\mb{k}\mb{q}}{\frac{\left(f_{c\mb{k}}-f_{d\mb{q}}\right){{\delta}_{\mb{k}\left(\mb{q}+\mb{\kappa}\right)}\left|{\mb{u}}^\ast_{c\mb{k}}\cdot {\mb{u}}_{d\mb{q}}\right|}^2}{E_{c\mb{k}}-E_{d\mb{q}}-\hbar \omega -i\hbar {\gamma}_{c\mb{k}d\mb{q}}}},                                        
\end{equation} 
\begin{equation} \label{eq_61_} 
{\sigma}^{\left(st\right)}\left(\omega ,\mb{\kappa}\right)=\frac{e^2g}{4{\pi}^2{\kappa}^2}\sum_{cd\mb{k}\mb{q}}{\left(i\omega -{\gamma}_{c\mb{k}d\mb{q}}\right)\frac{\left(f_{c\mb{k}}-f_{d\mb{q}}\right){{\delta}_{\mb{k}\left(\mb{q}+\mb{\kappa}\right)}\left|{\mb{u}}^\ast_{c\mb{k}}\cdot {\mb{u}}_{d\mb{q}}\right|}^2}{ E_{c\mb{k}}-E_{d\mb{q}}-\hbar \omega -i\hbar {\gamma}_{c\mb{k}d\mb{q}}}}.                     
\end{equation} 

As an important example, we compare the values of $ {\sigma}^{\left(st\right)}\left(0,\mb{\kappa}\right) $ and $ {\chi}^{\left(st\right)}\left(0,\mb{\kappa}\right) $. Using exactly the same approach as in the derivation of Eq.~\eqref{eq_57_}, we obtain $ \mathrm{Re}[{\sigma}^{\left(st\right)}\left(0,\mb{\kappa}\right)] \neq 0 $, but $ {\mathrm{Im}[\chi}^{\left(st\right)}\left(0,\mb{\kappa}\right)] =0 $. The relaxation operator \eqref{eq_25_} corresponds to the susceptibility given by Eq.~\eqref{eq_58_}, which leads to $ \mathrm{Im}[\chi \left(0,\mb{\kappa}\right)]={\mathrm{Im}[\chi}^{\left(st\right)}\left(0,\mb{\kappa}\right)] = 0 $. This suggests that, if one wants to use the standard relaxation operator Eq.~\eqref{eq_2_} for low frequencies and finite values of $ \mb{\kappa} $, one can get more adequate results from calculating the susceptibility $ \chi $ rather than the conductivity, since it is the condition $ \mathrm{Im}\chi \left(0,\mb{\kappa}\right)=0 $ that corresponds to the correct stationary state for finite values of $ \mb{\kappa} $. Therefore, below we compare the susceptibilities derived with different relaxation operators rather than conductivities.

\subsection{Comparison between the standard and new model of the relaxation operator for graphene}

Consider monolayer graphene, for which the wave functions $ {\mb{u}}_{c\mb{k}} $ and electron energy dispersion $ E_{c\mb{k}} $ are given in Appendix B (see  Eqs.~\eqref{appendixB_eq_8}, \eqref{appendixB_eq_9}). 
If we assume that the relaxation rate is a constant, $\gamma_{c\mb{k}d\mb{q}} = \gamma$, then the susceptibility given in Eq.~\eqref{eq_58_} is written as
\begin{align}
\chi(\omega,\mb{\kappa}) = - \frac{e^2 g}{4\pi^2\kappa^2} \sum_{cd} \int d^2\mb{q} \frac{ (f_{c,\mb{q}+\mb{\kappa}} - f_{d\mb{q}}) \left| u^\ast_{c,\mb{q}+\mb{\kappa}} \cdot u_{d\mb{q}}\right|^2 (E_{c,\mb{q}+\mb{\kappa}}-E_{d\mb{q}}) }
{ (E_{c,\mb{q}+\mb{\kappa}}-E_{d\mb{q}})^2 - \hbar^2\omega^2 - 2i\hbar^2\omega \gamma } .
\label{eq63} 
\end{align}
The above expression can be calculated numerically. However, when $\omega \to 0$, the denominator can become zero (even for $ \mb{\kappa}\neq 0 $ if $ c=d $), and the numerical integration does not work. So, we need to analyze the behavior of the susceptibility in the vicinity of $\omega = 0$. The denominator of $\chi(\omega,\mb{\kappa})$ can be written as
\begin{align}
\frac{1}{ (E_{c,\mb{q}+\mb{\kappa}}-E_{d\mb{q}})^2 - \hbar^2\omega^2 - 2i\hbar^2\omega \gamma }  
=
\frac{ (E_{c,\mb{q}+\mb{\kappa}}-E_{d\mb{q}})^2 - \hbar^2\omega^2 + 2i\hbar^2\omega \gamma }
{ \left( (E_{c,\mb{q}+\mb{\kappa}}-E_{d\mb{q}})^2 - \hbar^2\omega^2 \right)^2 + 4 \hbar^4 \omega^2 \gamma^2 }  .
\end{align}
Therefore, 
\begin{align}
&\phantom{{}={}}\lim\limits_{\omega\rightarrow 0} \frac{1}{ (E_{c,\mb{q}+\mb{\kappa}}-E_{d\mb{q}})^2 - \hbar^2\omega^2 - 2i\hbar^2\omega \gamma }   \nonumber \\
&= 
\mathrm{V.p.} \left\{ \frac{1}{ (E_{c,\mb{q}+\mb{\kappa}}-E_{d\mb{q}})^2 } \right\}
+
i\pi \delta( (E_{c,\mb{q}+\mb{\kappa}}-E_{d\mb{q}})^2 )  ,
\end{align}
where $\mathrm{V.p.}$ stands for the principal value of the integral. As a result, the susceptibility in the zero-frequency limit is given by 
\begin{align}
&\lim\limits_{\omega\rightarrow 0} \chi(\omega,\mb{\kappa})  
= - \frac{e^2 g}{4\pi^2\kappa^2} \sum_{cd} \int d^2\mb{q} 
\mathrm{V.p.} \left\{
\frac{ (f_{c,\mb{q}+\mb{\kappa}} - f_{d\mb{q}}) \left| u^\ast_{c,\mb{q}+\mb{\kappa}} \cdot u_{d\mb{q}}\right|^2 }
{ E_{c,\mb{q}+\mb{\kappa}}-E_{d\mb{q}} } \right\}  \nonumber \\
&- i \pi \frac{e^2 g}{4\pi^2\kappa^2} \sum_{cd} \int d^2\mb{q} (f_{c,\mb{q}+\mb{\kappa}} - f_{d\mb{q}}) \left| u^\ast_{c,\mb{q}+\mb{\kappa}} \cdot u_{d\mb{q}}\right|^2 (E_{c,\mb{q}+\mb{\kappa}}-E_{d\mb{q}}) \delta( (E_{c,\mb{q}+\mb{\kappa}}-E_{d\mb{q}})^2 )   \nonumber \\
&= - \frac{e^2 g}{4\pi^2\kappa^2} \sum_{cd} \int d^2\mb{q} 
\mathrm{V.p.} \left\{
\frac{ (f_{c,\mb{q}+\mb{\kappa}} - f_{d\mb{q}}) \left| u^\ast_{c,\mb{q}+\mb{\kappa}} \cdot u_{d\mb{q}}\right|^2 }
{ E_{c,\mb{q}+\mb{\kappa}}-E_{d\mb{q}} }  \right\}  \nonumber \\
&- i \pi \frac{e^2 g}{8\pi^2\kappa^2} \sum_{cd} \int d^2\mb{q} (f_{c,\mb{q}+\mb{\kappa}} - f_{d\mb{q}}) \left| u^\ast_{c,\mb{q}+\mb{\kappa}} \cdot u_{d\mb{q}}\right|^2 \delta( E_{c,\mb{q}+\mb{\kappa}}-E_{d\mb{q}} ) .
\end{align}
One can see that the imaginary part of $\chi(\omega,\mb{\kappa})$ is zero when $\omega\rightarrow 0$  for $ \mb{\kappa}\neq 0 $. 

Now we calculate the susceptibility of intrinsic monolayer graphene by carrying out the integration in $k$-space numerically. In the plots below, we assumed $T=300~\mathrm{K}$ and $\gamma = 10^{14}$  s$^{-1}$ $ \simeq 16$ THz. 

\begin{figure}[htb]
	\centering
	\begin{subfigure}{0.45\textwidth}
		\centering
		\includegraphics[width=\linewidth]{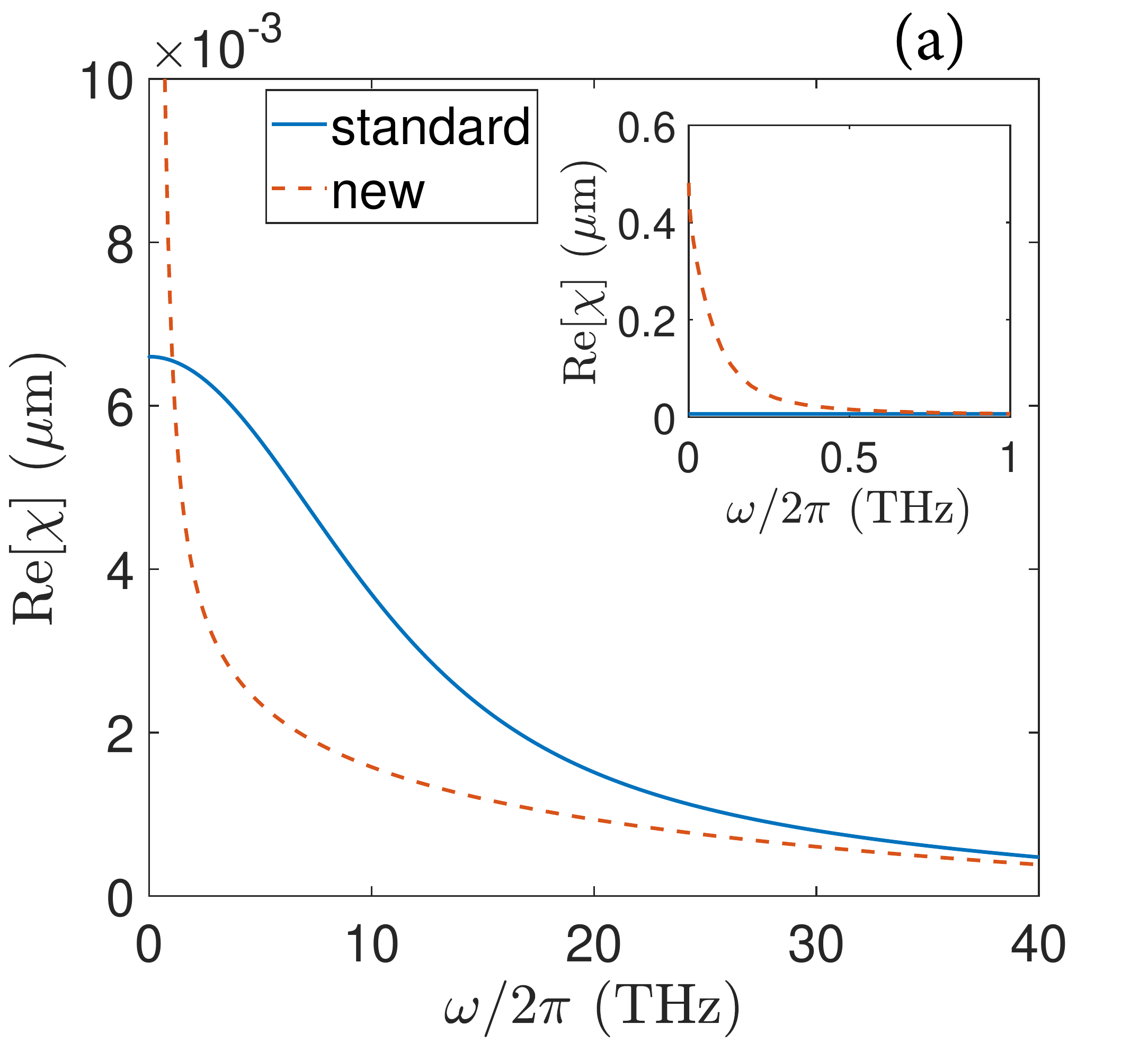}
	\end{subfigure}
	\begin{subfigure}{0.45\textwidth}
		\includegraphics[width=\linewidth]{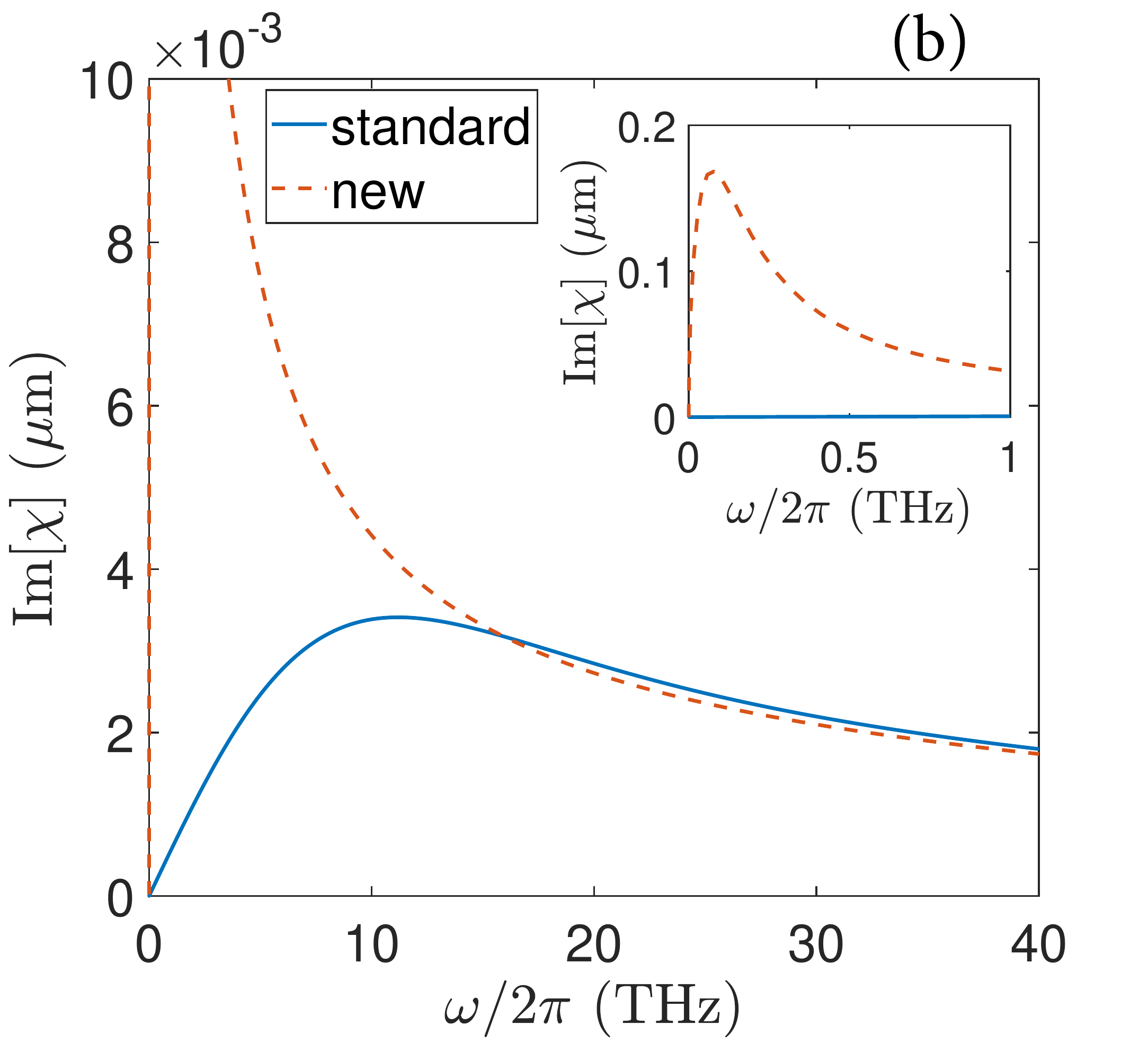}
	\end{subfigure}
	\caption{ The real part (a) and imaginary part (b) of the surface susceptibility for undoped monolayer graphene as a function of frequency for the two models: standard (solid blue curve) and new (dashed red curve). The plots are calculated for $T=300~\mathrm{K}$, $\gamma = 10^{14}$  s$^{-1}$ $ \simeq 16$ THz, and $\kappa/2\pi = 2~\mu\mathrm{m}^{-1}$. The inset in each figure shows the curves near zero frequency. }
	\label{Fig:chi_vs_w}
\end{figure}

In Fig.~\ref{Fig:chi_vs_w} we show the susceptibility as a function of frequency, calculated with the new model Eq.~(\ref{eq63})  in comparison with the standard model, Eq.~(\ref{eq_60_}). The inset shows the behavior near $\omega=0$. The difference between the predictions of the two models is very large when $\omega \leq \gamma$. In the high frequency limit $\omega \gg \gamma$ the models give a very similar result. 

%%%%%%%%%%%%%%%%%%%%%%%%%%%%%%%%%%%%

%%%%%%%%%%%%%%%%%%%%%%%%%%%%%%%%%%%%

\begin{figure}[htb]
	\centering
	\begin{subfigure}{0.45\textwidth}
		\centering
		\includegraphics[width=\linewidth]{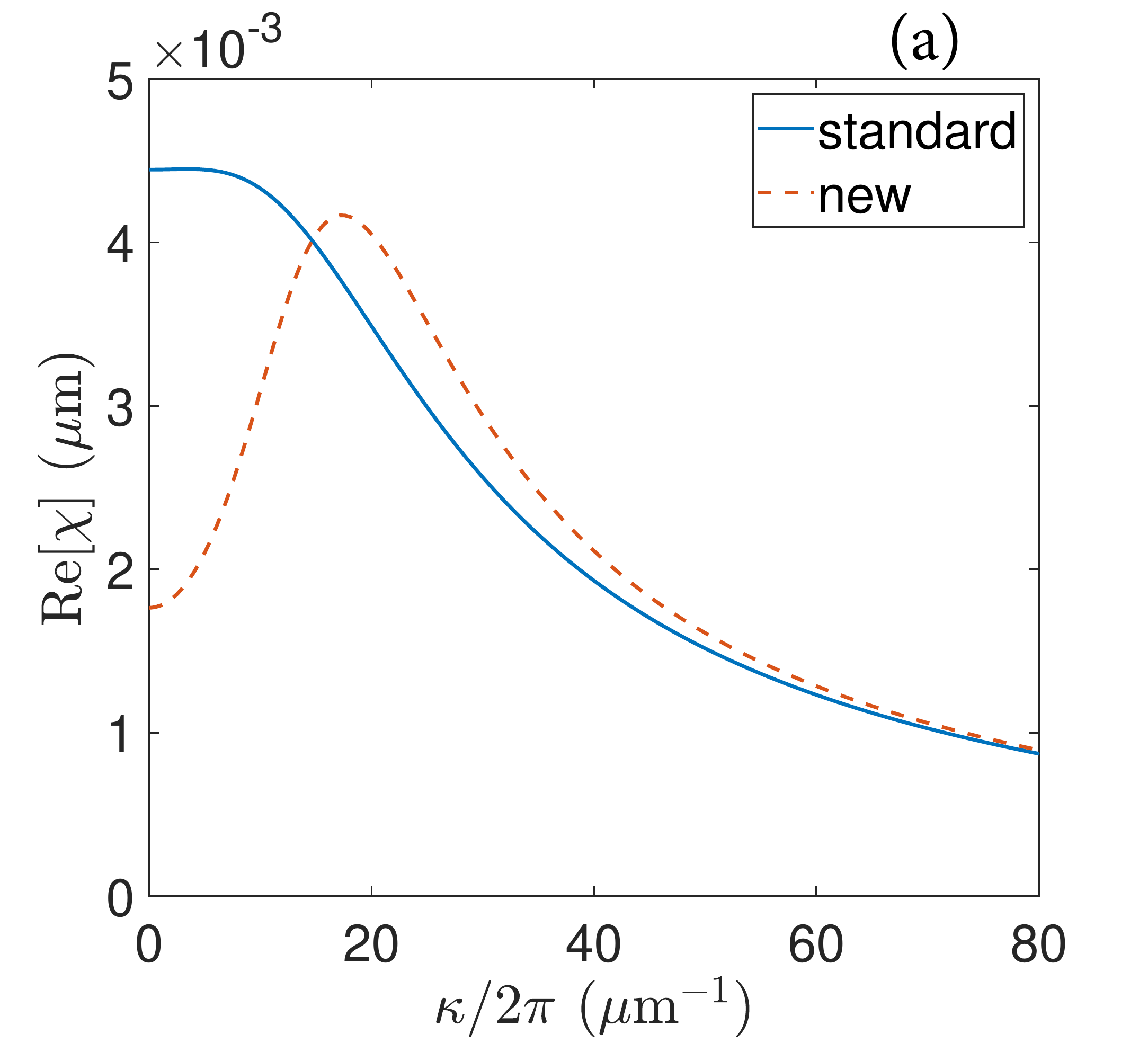}
	\end{subfigure}
	\begin{subfigure}{0.45\textwidth}
		\includegraphics[width=\linewidth]{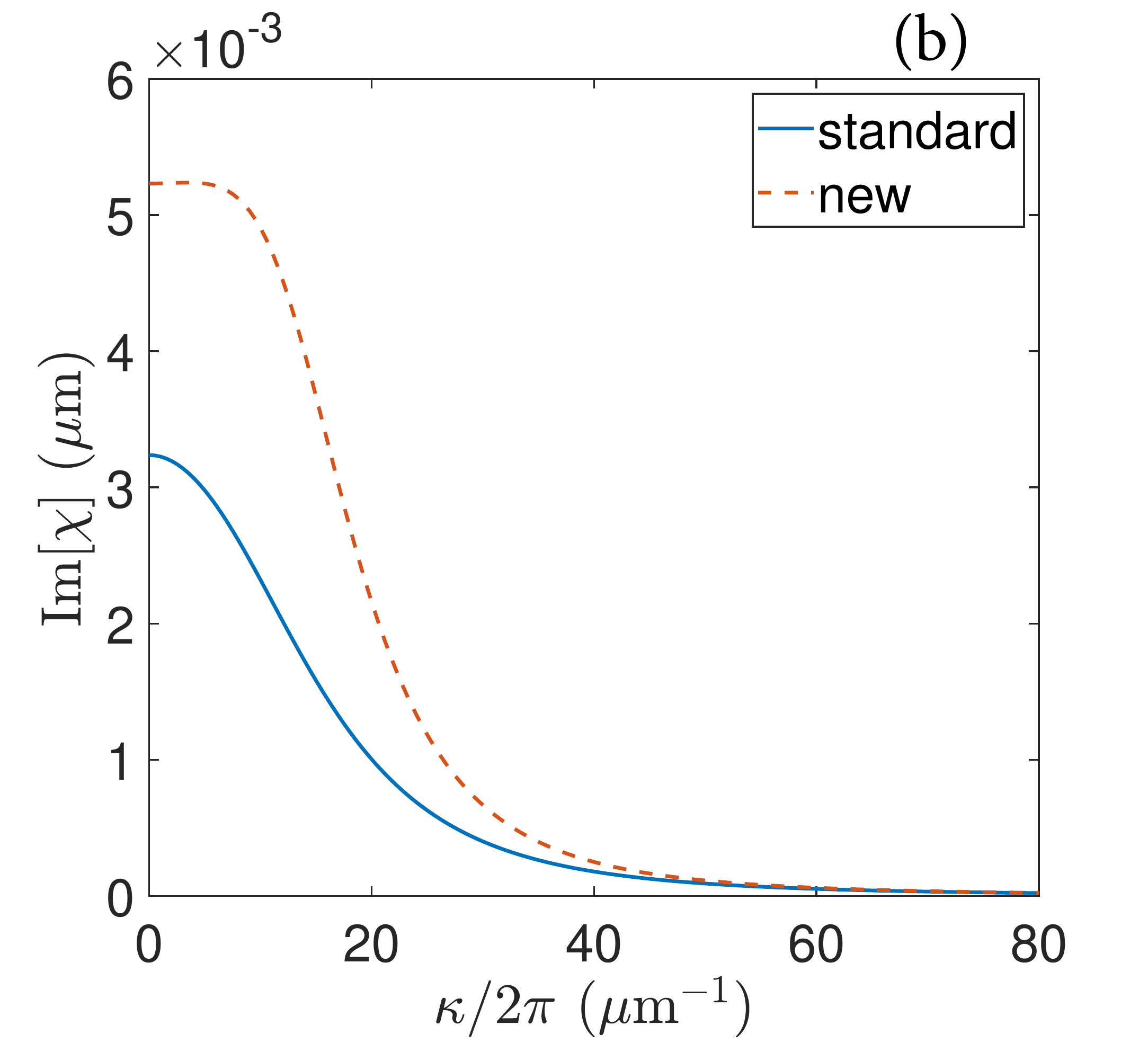}
	\end{subfigure}
	\caption{ The real part (a) and imaginary part (b) of the surface susceptibility for undoped monolayer graphene as a function of the wave vector $\kappa$ for the two models: standard (solid blue curve) and new (dashed red curve). The plots are calculated for $T=300~\mathrm{K}$ and $\gamma = 10^{14}$  s$^{-1}$ $ \simeq 16$ THz. The frequency is at $\omega=0.5\gamma$, namely $\omega/2\pi = 8.0~\mathrm{THz}$. }
	\label{Fig:chi_vs_kappa_w_eq_0.5_gamma}
\end{figure}

%%%%%%%%%%%%%%%%%%%%%%%%%%%%%%%%%%%%

\begin{figure}[htb]
	\centering
	\begin{subfigure}{0.45\textwidth}
		\centering
		\includegraphics[width=\linewidth]{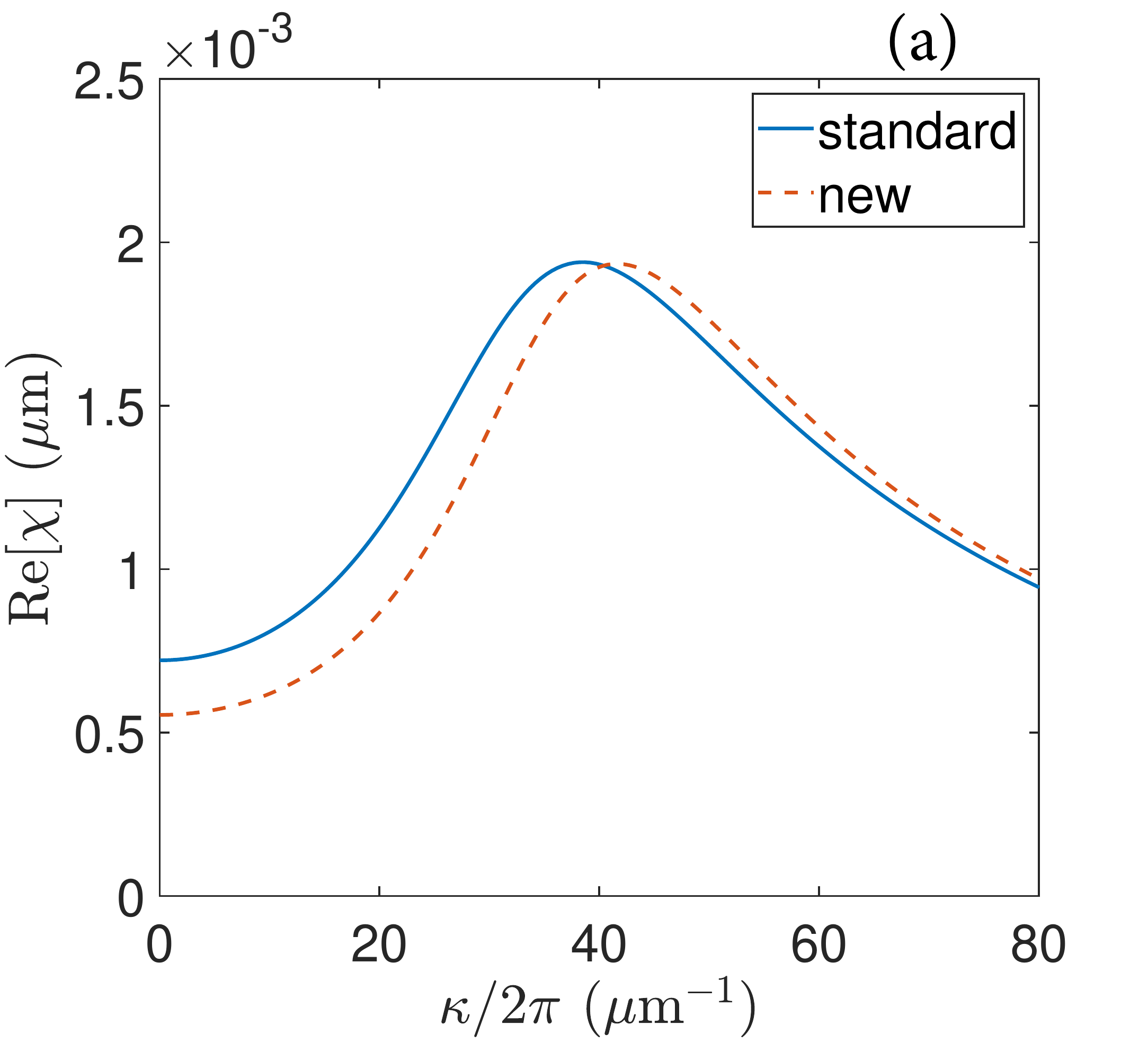}
	\end{subfigure}
	\begin{subfigure}{0.45\textwidth}
		\includegraphics[width=\linewidth]{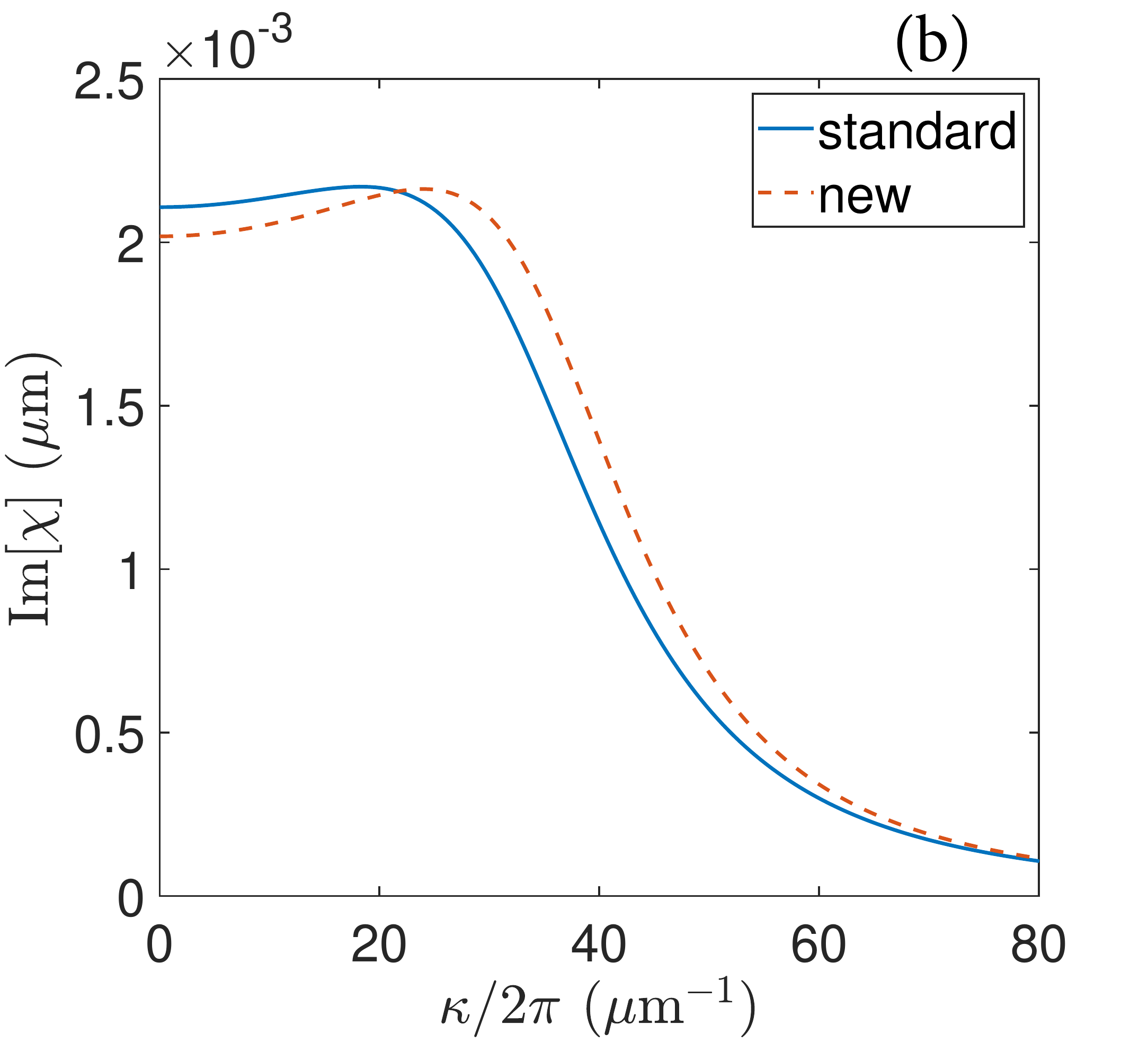}
	\end{subfigure}
	\caption{ The real part (a) and imaginary part (b) of the surface susceptibility for undoped monolayer graphene as a function of the wave vector $\kappa$ for the two models: standard (solid blue curve) and new (dashed red curve). The plots are calculated for $T=300~\mathrm{K}$ and $\gamma = 10^{14}$  s$^{-1}$ $ \simeq 16$ THz.  The frequency is at $\omega=2.0\gamma$, namely $\omega/2\pi = 31.8~\mathrm{THz}$. }
	\label{Fig:chi_vs_kappa_w_eq_2.0_gamma}
\end{figure}

Figure \ref{Fig:chi_vs_kappa_w_eq_0.5_gamma} shows the susceptibility as a function of the wave vector $\kappa$, while the frequency is fixed at $\omega = 0.5\gamma$. In Fig.~\ref{Fig:chi_vs_kappa_w_eq_2.0_gamma} we show the same dependence while the frequency is fixed at a higher value $\omega = 2.0\gamma$. Clearly, at low frequencies the models have a very different behavior when $\kappa$ approaches zero. The difference between the models becomes less important as the frequency gets larger than the relaxation rate.

\subsection{The case of $\kappa\rightarrow 0$ for graphene}
\label{sec:graphene_kappa_0}
In the case of ${\kappa} \rightarrow 0$, the susceptibility in the modified model is given in Eq.~\eqref{eq_59_}, and the susceptibility in the standard model is given in Eq.~\eqref{eq_60_}. Here we try to compare the results by finding the analytical expressions of the susceptibility. In order to do this, we consider the case of zero temperature, so the distribution of electrons is given by $f_{n\mb{k}} = \theta(E_F-E_{n\mb{k}})$, where $\theta(x)$ is the Heaviside function, and $E_F$ is the Fermi level, which is related to the Fermi wavevector $k_F$ by $E_F = \mathrm{sgn}(E_F) \hbar v_F k_F$, where $\mathrm{sgn}(x)$ is the sign function.

The detailed derivation is given in Appendix \ref{appendix:chi_kappa_0_derivation}. For the modified model, we find the following expression for part of $\chi(\omega,0)$ due to the interband transitions,  
\begin{align}
\chi_{\mathrm{inter}}(\omega, 0) 
&=
-\frac{e^2 g}{16\pi}
\frac{1}{\hbar\omega\sqrt{\left( 1+2i\gamma/\omega \right) }}
\ln
\left[
\frac{2 v_F k_F - \omega\sqrt{\left( 1+2i\gamma/\omega \right)} }{2 v_F k_F + \omega\sqrt{\left( 1+2i\gamma/\omega \right)} }
\right]  ,
\end{align}
where the branch of the square root should have $\mathrm{Re}[\sqrt{\left( 1+2i\gamma/\omega \right) }] > 0$.
The contribution of intraband transitions is found to be
\begin{align}
\chi_{\mathrm{intra}}(\omega, 0) 
&=
-\frac{e^2 g}{4\pi\hbar} 
\frac{v_F k_F}{\omega^2 (1+2i\gamma/\omega) } .
\end{align}

For the standard relaxation operator, the contribution of interband transitions is 
\begin{align}
\chi^{\mathrm{(st)}}_{\mathrm{inter}}(\omega, 0) 
&=
-\frac{e^2 g}{16\pi}
\frac{1}{\hbar\omega (1+i\gamma/\omega) }
\ln
\left[
\frac{2 v_F k_F - \omega (1+i\gamma/\omega) } {2 v_F k_F + \omega (1+i\gamma/\omega) }
\right],
\end{align}
and the contribution of the intraband transitions is
\begin{align}
{\chi}^{\mathrm{(st)}}_{\mathrm{intra}} \left(\omega, 0\right)
&=
-\frac{e^2g}{4\pi\hbar} \frac{v_F k_F}{\omega^2(1 + i\gamma/\omega)^2} .
\end{align}

These results show that the functions $\chi(\omega,0)$ calculated in the standard and modified models tend to be the same at large frequencies $\omega \gg \gamma$, while they are quite different in the region where $\omega$ is of the order of or smaller than $\gamma$.

\section{Conclusions}

In conclusion, we derived the phenomenological relaxation operator for quasiparticles in a crystalline solid, which has a number of important advantages as compared to widely used models. Our relaxation operator  is valid for charged carriers in solids with an arbitrary energy dispersion, in particular the Dirac spectrum; it preserves the continuity equation while including both intraband and interband transitions; it allows one to obtain both the stationary ``current-free'' regime in equilibrium and the well-known ohmic direct current  regime in the limit of a uniform static field; and it is  much  simpler and more general than the model proposed in \cite{mermin1970}.

 We demonstrated a significant difference between the results of applying the standard and modified models of relaxation of quantum coherence in the low-frequency region. We believe that the proposed relaxation operator model should be used in a wide range of problems related to the interaction of electromagnetic fields with condensed matter systems.

\begin{acknowledgments}
This work has been supported in part by the Air Force Office for Scientific Research 
Grant No.~FA9550-17-1-0341, National Science Foundation Award No.~1936276, and Texas A\&M University through X-grant and T3-grant programs.
M.T. acknowledges the support from RFBR Grant No. 18-29-19091mk. 
M.E. acknowledges the support from Federal Research Center Institute of Applied Physics of the Russian Academy of Sciences
(Project No. 0035-2019-004).

\end{acknowledgments}

\appendix

\section{Relaxation matrix for a TRS-invariant Hamiltonian}
\label{sec_appendix_TRS}

Consider the Schr\"{o}dinger equation $ i\hbar \dot{\mb{\mathrm{U}}}={\hat{H}}_0\mb{\mathrm{U}} $, where $ \mb{\mathrm{U}} $ is an \textit{N}-component vector. The solution $ \mb{\mathrm{U}}\left(t\right) $ is related to the time-reversed solution $ \tilde{\mb{\mathrm{U}}}\left(t\right) $ by $ \tilde{\mb{\mathrm{U}}}(t)=\hat{Q}{\mb{\mathrm{U}}}^{\mb{*}}(-t) $, where $ \hat{Q} $ is a certain linear operator which does not have to be specified. Invariance with respect to time reversal means that vectors $ \tilde{\mb{\mathrm{U}}}(t) $ and $ \mb{\mathrm{U}}(t) $ must satisfy the Schr\"{o}dinger equation with the same Hamiltonian $ {\hat{H}}_0 $, which implies: 
\begin{equation} \label{appendixA_eq_1}
{\hat{H}}^\ast_0={\hat{Q}}^{-1}{\hat{H}}_0\hat{Q}.                                  \end{equation}
If $ {\mb{\mathrm{U}}}_{c\mb{k}}\left(\mb{r}\right)={\mb{u}}_{c\mb{k}}{\mathrm{e}}^{i\mb{kr}} $ is the eigenvector of the Hamiltonian $ {\hat{H}}_0 $ for energy $ E_{c\mb{k}} $, then another eigenvector for the state with the same energy corresponds to the time reversal operation,
\begin{equation} \label{appendixA_eq_2}
{\mb{u}}_{c-\mb{k}}={\mathrm{e}}^{i{\varphi}_{c\mb{k}}}\hat{Q}{\mb{u}}^{\mb{*}}_{c\mb{k}}.                                     \end{equation}
It follows from Eq.~\eqref{appendixA_eq_1} and the hermiticity of the operator $ {\hat{H}}_0 $ that
\begin{equation} \label{appendixA_eq_3}
{\hat{Q}}^{-1}=\pm {\hat{Q}}^\ast, ~~~ {\hat{Q}}^{-1}=\pm {\hat{Q}}^{\dagger}.
\end{equation}
Using Eq.~\eqref{appendixA_eq_2} to transform the scalar product in Eq.~\eqref{eq_35_}, we get 
\[\left({\mb{u}}^\ast_{c-\mb{k}}\cdot {\mb{u}}_{d-\mb{q}}\right)={\mathrm{e}}^{i\left({\varphi}_{d\mb{q}}-{\varphi}_{c\mb{k}}\right)}\left(\hat{Q}{\mb{u}}^{\mb{*}}_{d\mb{q}}\cdot {\hat{Q}}^\ast{\mb{u}}_{c\mb{k}}\right)={\mathrm{e}}^{i\left({\varphi}_{d\mb{q}}-{\varphi}_{c\mb{k}}\right)}\left({\mb{u}}^{\mb{*}}_{d\mb{q}}{\cdot \hat{Q}{\hat{Q}}^{\dagger}\mb{u}}_{c\mb{k}}\right).\] 
Equation \eqref{appendixA_eq_3} means that $ {\hat{Q}\hat{Q}}^{\dagger}=\pm \hat{1} $, where $ \hat{1} $ is the identity matrix. Using this,  we arrive at the following expression for the coefficients $ G_{c\mb{k}d\mb{q}} $, 
\begin{equation} \label{appendixA_eq_4}
G_{c\mb{k}d\mb{q}}= \xi e^{i\left({\varphi}_{d\mb{q}}-{\varphi}_{c\mb{k}}\right)},
\end{equation}        
where $ \xi =\pm 1 $ for $ {\hat{Q}\hat{Q}}^{\dagger}=\pm \hat{1} $. Equation \eqref{eq_36_} follows from Eq.~\eqref{appendixA_eq_4}.

\section{Relaxation operator for quasiparticles in monolayer graphene}
\label{appendix_graphene}

%\label{subsec_appendix_graphene}

The Bloch functions for quasiparticles in monolayer graphene are given by \cite{katsnelson2012graphene}
\begin{equation} \label{appendixB_eq_1}
\langle \mb{r} | \alpha \rangle ={\mathrm{\Psi}}_{s\mb{k}}\left(\mb{r}\right)={\mathrm{e}}^{i\mb{kr}}\left[{\lambda}_A{\psi}^A_{\mb{k}}\left(\mb{r}\right)+{\lambda}_B{\psi}^B_{\mb{k}}\left(\mb{r}\right)\right],                                             \end{equation}
where
\begin{equation} \label{appendixB_eq_2}
{\psi}^A_{\mb{k}}\left(\mb{r}\right)=\sum_{{\mb{r}}_A}{{\mathrm{e}}^{i\mb{k}\left({\mb{r}}_A-\mb{r}\right)}}X\left(\mb{r}-{\mb{r}}_A\right),~~~ {\psi}^B_{\mb{k}}\left(\mb{r}\right)=\sum_{{\mb{r}}_B}{{\mathrm{e}}^{i\mb{k}\left({\mb{r}}_B-\mb{r}\right)}}X\left(\mb{r}-{\mb{r}}_B\right) ,
\end{equation}
 $ {\mb{r}}_{A,B} $ are atom positions in two sublattices, and $ X\left(\mb{r}-{\mb{r}}_{A,B}\right) $ is the Wannier function. The set of vectors $ \left( \begin{array}{c}
{\lambda}_A \\ 
{\lambda}_B \end{array}
\right) $ and the energies $ E $ of the quasiparticles  correspond to eigenvectors and eigenvalues of the transformation which in the tight binding  approximation has the following form \cite{katsnelson2012graphene},
\begin{equation} \label{appendixB_eq_3}
\hbar \nu \left( \begin{array}{cc}
0 & S\left(\mb{k}\right) \\ 
S^\ast\left(\mb{k}\right) & 0 \end{array}
\right)\cdot \left( \begin{array}{c}
{\lambda}_A \\ 
{\lambda}_B \end{array}
\right)=E\left( \begin{array}{c}
{\lambda}_A \\ 
{\lambda}_B \end{array}
\right) , 
\end{equation}
where
\begin{equation} \label{appendixB_eq_4}
S\left(\mb{k}\right)=2\mathrm{exp}\left(\frac{ik_xa}{2}\right)\mathrm{cos}\left(\frac{\sqrt{3}k_ya}{2}\right)+\mathrm{exp}\left(-ik_xa\right) ,
\end{equation} 
 and $ \nu $ is a normalized hopping parameter. From Eq.~\eqref{appendixB_eq_3} we obtain

\begin{equation} \label{appendixB_eq_5}
E_{c\mb{k}}=c\hbar \nu  \left|S\left(\mb{k}\right)\right|,~~~   
{{\lambda}_A}/{{\lambda}_B={cS\left(\mb{k}\right)}/{\left|S\left(\mb{k}\right)\right|}} , 
\end{equation}
where $ c=\pm 1 $  is the band index. The relations \eqref{appendixB_eq_1}, \eqref{appendixB_eq_2}, \eqref{appendixB_eq_4} and the second of Eqs.~\eqref{appendixB_eq_5} define the set of wave functions $ {\mathrm{\Psi}}_{s\mb{k}}\left(\mb{r}\right) $. The spin-orbit coupling which leads to the spin dependence of energy is negligible in this case.

Taking into account  that  $ S\left(-\mb{k}\right)=S^\ast\left(\mb{k}\right) $, we get $ E_{c\mb{k}}=E_{c-\mb{k}} $, $ {\mathrm{\Psi}}_{c-\mb{k}}\left(\mb{r}\right)={\mathrm{\Psi}}^\ast_{c\mb{k}}\left(\mb{r}\right) $, which should be expected from the TRS property of the system. The last equation indicates that one can use the relaxation operator in the form of Eq.~\eqref{eq_22_}. Note, however that the relation  $ S\left(-\mb{k}\right)=S^\ast\left(\mb{k}\right) $ refers to the complete Brillouin zone that includes two Dirac points: $ \mb{K}=\left(\frac{2\pi}{3a},\frac{2\pi}{3\sqrt{3}a}\right) $ and $ {\mb{K}}^{\mb{'}}=\left(\frac{2\pi}{3a}\mb{,-}\frac{2\pi}{3\sqrt{3}a}\right) $. Therefore, the replacement $ -\mb{k} \leftrightarrow \mb{k} $ can connect quasiparticles from different valleys and the intervalley scattering can contribute to the relaxation rate \cite{gantmakher1987carrier}.

In the vicinity of the Dirac points, it is convenient to use approximate expressions \cite{katsnelson2012graphene}
\begin{equation} \label{appendixB_eq_6}
S\left(\mb{k}\right)\propto \left(\delta k_x-{i\delta k}_y\right) , 
\end{equation}
where $ \delta \mb{k}=\mb{k}-\mb{K} $ or $ \delta \mb{k}=\mb{k}-{\mb{K}}^{\mb{'}} $. Equation \eqref{appendixB_eq_6} leads to an effective Hamiltonian of the type \eqref{eq_29_}, in which $ \frac{{\partial}^2\hat{E}}{\partial k_i\partial k_j}\to 0 $:
\begin{equation} \label{appendixB_eq_7}
{\hat{H}}_0=v_F\hat{\mb{p}}\hat{\mb{\sigma}} ,
\end{equation}
where $ \hat{\mb{\sigma}}={\mb{x}}_0{\hat{\sigma}}_x+{\mb{y}}_0{\hat{\sigma}}_y $ and $ v_F=\frac{3a\nu}{2} $ is the Fermi velocity. The Hamiltonian \eqref{appendixB_eq_7} corresponds to the energy dispersion for massless Dirac quasiparticles given by 
\begin{equation} \label{appendixB_eq_8}
E_{c\mb{k}}=c\hbar v_Fk ,
\end{equation}
and the eigenfunctions 
\begin{equation} \label{appendixB_eq_9}
{\mb{\mathrm{U}}}_{c\mb{k}}\left(\mb{r}\right)={\mathrm{e}}^{i\mb{kr}}{\mb{u}}_{c\mb{k}},~~~ {\mb{u}}_{c\mb{k}}=\frac{1}{\sqrt{2}}\left( \begin{array}{c}
c \\ 
{\mathrm{e}}^{i\theta \left(\mb{k}\right)} \end{array}
\right) ,
\end{equation}
where the indices $ c=\pm 1 $ denote the conduction and valence bands, respectively, $ \theta \left(\mb{k}\right) $ is the angle between the quasimomentum $ \hbar \mb{k} $ and the \textit{x} axis in the plane of the monolayer, the vector $\mb{k}$ is defined in the \textit{k}-space relative to the point $ \mb{K} $ or $ {\mb{K}}^{\mb{'}} $; the degeneracy over the valley index is assumed. 

Note that the model Hamiltonian \eqref{appendixB_eq_7} is TRS invariant so that Eq.~\eqref{appendixA_eq_1} with operator $ \hat{Q}=i{\hat{\sigma}}_y $ is satisfied and the eigenvectors \eqref{appendixB_eq_9} satisfy the relation $ {\mb{u}}_{c -\mb{k}}=e^{{i\varphi}_{c\mb{k}}}\hat{Q}{\mb{u}}^{\mb{*}}_{c\mb{k}} $, where
\[{\varphi}_{c\mb{k}}=\left\{ \begin{array}{c}
\theta \left(\mb{k}\right),~ c=1 \\ 
\theta \left(\mb{k}\right)+\pi ,~  c=-1 \end{array}
\right..\] 
Therefore, one can use Eq.~\eqref{appendixA_eq_4} to find the coefficients $ G_{c\mb{k}d\mb{q}} $ which determine the form of the relaxation operator in Eq.~\eqref{eq_25_}. As a result, since $ \hat{Q}{\hat{Q}}^{\dagger}=\hat{1} $, we get the following relaxation operator for carriers in graphene in the vicinity of a Dirac point,
\begin{equation} \label{appendixB_eq_10}
R_{c\mb{k}d\mb{q}}=-{\gamma}_{c\mb{k}d\mb{q}}\left({\rho}_{c\mb{k}d\mb{q}}-cd{\mathrm{e}}^{i\left[\theta \left(\mb{q}\right)-\theta \left(\mb{k}\right)\right]}{\rho}_{d\,-\mb{q}\,c\,-\mb{k}}\right) .
\end{equation}
One can obtain the same expression by substituting Eq.~\eqref{appendixB_eq_9} for the eigenfunctions  directly into Eqs.~\eqref{eq_25_} and \eqref{eq_35_}.

\section{Relaxation operator for quasiparticles in a TRS-breaking Weyl semimetal}
\label{appendix_WSM}

We use the ``minimal'' model Hamiltonian for a TRS-breaking Weyl semimetal from \cite{chen2019}, which describes the two bands touching in two separated Weyl nodes in the quasimomentum space, near which electrons have a three-dimensional Dirac spectrum: 
\begin{equation} \label{appendixB_eq_11}
{\hat{H}}_0=v_F\left(\frac{{\hat{P}}^2-{\hbar}^2m}{2\hbar b}{\hat{\sigma}}_x+{\hat{p}}_y{\hat{\sigma}}_y+{\hat{p}}_z{\hat{\sigma}}_z\right) , 
\end{equation}
where $ {\hat{\sigma}}_{x,y,z} $ are the Pauli matrices and the operator $ {\hat{P}}^2 $ can be defined in one of three ways: (1) $ {\hat{P}}^2={\hat{p}}^2_x $, (2) $ {\hat{P}}^2={\hat{p}}^2_x+{\hat{p}}^2_y $,  or (3) $ {\hat{P}}^2={\hat{p}}^2_x+{\hat{p}}^2_y+{\hat{p}}^2_z $.

The Hamiltonian in Eq.~\eqref{appendixB_eq_11} is not TRS invariant, which gives rise, in particular, to the gyrotropy and the anomalous Hall effect in the absence of an external magnetic field \cite{chen2019, chen2019-2}. Physically, this corresponds to a material with some kind of a magnetic order. 

The bulk states of the Hamiltonian \eqref{appendixB_eq_11} have the energy dispersion  
\begin{equation} \label{appendixB_eq_12}
E_{c\mb{k}}=c\hbar v_F\sqrt{K^2_x+k^2_y+k^2_z} ,
\end{equation}
and eigenvectors
\begin{equation} \label{appendixB_eq_13}
{\mb{\mathrm{U}}}_{c\mb{k}}\left(\mb{r}\right)={\mathrm{e}}^{i\mb{kr}}{\mb{u}}_{c\mb{k}},~~~ {\mb{u}}_{c\mb{k}}=\frac{1}{\sqrt{2}}\left( \begin{array}{c}
\sqrt{1-c{\mathrm{cos} {\theta}_{\mb{k}}}}e^{-i{\phi}_{\mb{k}}} \\ 
c\sqrt{1+c{\mathrm{cos} {\theta}_{\mb{k}}}} \end{array}
\right) ,
\end{equation}
where $ {\mathrm{cos} {\theta}_{\mb{k}}}=\frac{k_y}{\sqrt{K^2_x+ {k}^2_y+{k}^2_z}} $, $ e^{i{\phi}_{\mb{k}}}=\frac{K_x+ik_z}{\sqrt{K^2_x+k^2_z}} $, and  (1) $ K_x=\frac{k^2_x-m}{2b} $, (2) $ K_x=\frac{k^2_x +{k}^2_y-m}{2b} $, or (3) $ K_x=\frac{k^2_x+{k}^2_y+{k}^2_z-m}{2b} $, respectively; $ c=\pm 1 $ (see \cite{chen2019}).

Due to the broken TRS in this system \cite{chen2019}, one cannot use the representation Eq.~\eqref{appendixA_eq_4} for the coefficients $ G_{c\mb{k}d\mb{q}} $, which determine the form of the relaxation operator Eq.~\eqref{eq_25_}. However, calculating these coefficients directly using Eq.~\eqref{eq_35_} with Eq.~\eqref{appendixB_eq_13} taken into account, we obtain 
\[G_{c\mb{k}d\mb{q}}=cde^{i{\phi}_{\mb{q}}-i{\phi}_{\mb{k}}}\frac{\sqrt{1+c{\mathrm{cos} {\theta}_{\mb{k}}}}\sqrt{1+d{\mathrm{cos} {\theta}_{\mb{q}}}}+cde^{i{\phi}_{\mb{q}}-i{\phi}_{\mb{k}}}\sqrt{1-c{\mathrm{cos} {\theta}_{\mb{k}}}}\sqrt{1-d{\mathrm{cos} {\theta}_{\mb{q}}}}}{\sqrt{1+c{\mathrm{cos} {\theta}_{\mb{k}}}}\sqrt{1+d{\mathrm{cos} {\theta}_{\mb{q}}}}+cd e^{-i{\phi}_{\mb{q}}+i{\phi}_{\mb{k}}}\sqrt{1-c{\mathrm{cos} {\theta}_{\mb{k}}}}\sqrt{1-d{\mathrm{cos} {\theta}_{\mb{q}}}}},\] 
whence it is easy to see that  $ |G_{c\mb{k}d\mb{q}}| = 1 $, despite the broken TRS of the system.

\section{The linear susceptibility for a uniform perturbation}
\label{sec_appendix_homogeneous}

From Eq.~\eqref{eq_52_} and taking into account  Eqs.~\eqref{eq_53_}, \eqref{eq_54_}, we obtain 
\begin{equation} \label{appendixC_eq_1}
{\tilde{\rho}}_{c\mb{k}d\mb{q}}=-e\frac{\left(E_{c\mb{k}}-E_{d\mb{q}}+\hbar \omega +i\hbar {\gamma}_{c\mb{k}d\mb{q}}\left(\mathrm{1}-{\left|G_{c\mb{k}d\mb{q}}\right|}^2\right)\right){\mathrm{\Phi}}_{\omega ;c\mb{k}d\mb{q}}\left(f_{c\mb{k}}-f_{d\mb{q}}\right)}{{\left(E_{c\mb{k}}-E_{d\mb{q}}\right)}^2-{\hbar}^2{\omega}^2-2{i\hbar}^2\omega {\gamma}_{c\mb{k}d\mb{q}}} .
\end{equation}
The expression for the linear susceptibility calculated by substituting Eq.~\eqref{appendixC_eq_1} into Eq.~\eqref{eq_44_} is 
\begin{equation}  \label{appendixC_eq_2}
\chi \left(\omega ,\mb{\kappa}\right)=-\frac{e^2g}{4{\pi}^2{\kappa}^2}\sum_{cd\mb{k}\mb{q}}{\frac{\left(f_{c\mb{k}}-f_{d\mb{q}}\right){{\delta}_{\mb{k}\left(\mb{q}+\mb{\kappa}\right)}\left|{\mb{u}}^\ast_{c\mb{k}}\cdot {\mb{u}}_{d\mb{q}}\right|}^2\left[E_{c\mb{k}}-E_{d\mb{q}}+\hbar \omega +i\hbar {\gamma}_{c\mb{k}d\mb{q}}\left(\mathrm{1}-{\left|G_{c\mb{k}d\mb{q}}\right|}^2\right)\right]}{{{\left(E_{c\mb{k}}-E_{d\mb{q}}\right)}^2-\hbar}^2{\omega}^2-2{i\hbar}^2\omega {\gamma}_{c\mb{k}d\mb{q}}}} .
\end{equation}
After rearranging the summation indices, taking into account the symmetry of the energy dispersion and populations with respect to the replacement $ \mb{k}\to -\mb{k} $, $ \mb{q}\to -\mb{q} $, we can convert Eq.~\eqref{appendixC_eq_2} into another form,
\begin{align}  \label{appendixC_eq_3}
\chi \left(\omega ,\mb{\kappa}\right)=-\frac{e^2}{4{\pi}^2{\kappa}^2}\sum_{cd\mb{k}\mb{q}}{\left[\frac{\left(f_{c\mb{k}}-f_{d\mb{q}}\right){\left(E_{c\mb{k}}-E_{d\mb{q}}\right)\left|{\mb{u}}^\ast_{c\mb{k}}\cdot {\mb{u}}_{d\mb{q}}\right|}^2   \frac{{\delta}_{\mb{k},\mb{q}+\mb{\kappa}}+{\delta}_{\mb{k},\mb{q}-\mb{\kappa}}}{2}}{{{\left(E_{c\mb{k}}-E_{d\mb{q}}\right)}^2-\hbar}^2{\omega}^2-2{i\hbar}^2\omega {\gamma}_{c\mb{k}\,d\mb{q}}}\right.} \nonumber \\ 
\left.+ \frac{\left(f_{c\mb{k}}-f_{d\mb{q}}\right){\left|{\mb{u}}^\ast_{c\mb{k}}\cdot {\mb{u}}_{d\mb{q}}\right|}^2\left(\hbar \omega +i\hbar {\gamma}_{c\mb{k}d\mb{q}}\left(\mathrm{1}-{\left|G_{c\mb{k}d\mb{q}}\right|}^2\right)\right)\frac{{\delta}_{\mb{q},\mb{k}-\mb{\kappa}}-{\delta}_{\mb{q},\mb{k}+\mb{\kappa}}}{2}}{{{\left(E_{c\mb{k}}-E_{d\mb{q}}\right)}^2-\hbar}^2{\omega}^2-2{i\hbar}^2\omega {\gamma}_{c\mb{k}\,d\mb{q}}}\right].        
\end{align}
In the limit of a homogeneous perturbing field, when $ {\kappa}/{k}\to 0 $ we obtain Eq.~\eqref{eq_59_}, which does not depend on the value of $ {\left|G_{c\mb{k}d\mb{q}}\right|}^2 $.

\section{Derivation of the linear susceptibility for graphene when $\kappa\rightarrow 0$}
\label{appendix:chi_kappa_0_derivation}

In this section, we show the derivation of $\chi(\omega, \mb{\kappa})$ for graphene in the limit of $|\mb{\kappa}| \to 0$, which we study in Section \ref{sec:graphene_kappa_0}. As stated in the main text, we consider the case of zero temperature, where the distribution of electrons is given by $f_{n\mb{k}} = \theta(E_F-E_{n\mb{k}})$, where $\theta(x)$ is the Heaviside function, and $E_F$ is the Fermi level, which is related to the Fermi wavevector $k_F$ by $E_F = \mathrm{sgn}(E_F) \hbar v_F k_F$, where $\mathrm{sgn}(x)$ is the sign function. 

In the case of ${\kappa} \rightarrow 0$, the susceptibility with the new relaxation operator is given by Eq.~\eqref{eq_59_}. Without loss of generality, we can choose $\mb{\kappa}$ in the $\hat{x}$ direction. For the contribution of interband transitions, using $x_{c\mb{k}v\mb{k}} = \sin\theta(\mb{k})/2k$, we find that the susceptibility can be written as 
\begin{align}
\chi_{\mathrm{inter}}(\omega, 0) 
&= 
- \frac{e^2 g}{4\pi^2} \int d^2 \mb{k} \left[ 2 \frac{\sin^2\theta(\mb{k})}{4k^2} \frac{(f_{c\mb{k}}-f_{v\mb{k}})2\hbar v_F k} {(2\hbar v_F k)^2 - \hbar^2\omega^2 - 2i\hbar^2\omega \gamma} \right] \nonumber \\
&= 
- \frac{e^2 g}{4\pi^2} \int_0^\infty k d k \int_0^{2\pi} d\theta \left[ \frac{\sin^2\theta}{k} \frac{(f_{ck}-f_{vk}) \hbar v_F} {(2\hbar v_F k)^2 - \hbar^2\omega^2 - 2i\hbar^2\omega \gamma} \right] \nonumber \\
&= 
- \frac{e^2 g}{4\pi} \int_0^\infty d k \left[ \frac{(f_{ck}-f_{vk}) \hbar v_F} {(2\hbar v_F k)^2 - \hbar^2\omega^2 - 2i\hbar^2\omega \gamma} \right] \nonumber \\
&= 
\frac{e^2 g}{4\pi} \int_{k_F}^\infty d k \left[ \frac{ \hbar v_F} {(2\hbar v_F k)^2 - \hbar^2\omega^2 - 2i\hbar^2\omega \gamma} \right] \nonumber \\
&=
-\frac{e^2 g}{16\pi}
\frac{1}{\hbar\omega\sqrt{\left( 1+2i\gamma/\omega \right) }}
\ln
\left[
\frac{2 v_F k_F - \omega\sqrt{\left( 1+2i\gamma/\omega \right)} }{2 v_F k_F + \omega\sqrt{\left( 1+2i\gamma/\omega \right)} }
\right]  ,
\end{align}
where the branch of the square root should have $\mathrm{Re}[\sqrt{\left( 1+2i\gamma/\omega \right) }] > 0$.

The contribution of the intraband transitions is 
\begin{align}
\chi_{\mathrm{intra}}(\omega, 0) 
&= \frac{e^2 g}{4\pi^2} \int d^2 \mb{k} \sum_{n=c,v}
{\frac{\left(\mb{n}\frac{\partial}{\partial \mb{k}}E_{n\mb{k}}\right)\left(\mb{n}\frac{\partial}{\partial \mb{k}}f_{n\mb{k}}\right)  }{{\hbar}^2\omega^2+2{i\hbar}^2\omega \gamma}}
\nonumber \\
&=
\frac{e^2 g}{4\pi^2} 
\frac{1}{\hbar^2\omega^2+2{i\hbar}^2\omega \gamma}
\int d^2 \mb{k}
 \sum_{n=c,v}
\left(\mb{n}\frac{\partial}{\partial \mb{k}}E_{n\mb{k}}\right)^2 
\frac{\partial}{\partial E_{n\mb{k}} } f_{n\mb{k}}   \nonumber \\
&=
\frac{e^2 g}{4\pi^2} 
\frac{v_F^2}{\omega^2+2i \omega \gamma}
\int d^2 \mb{k}
\sum_{n=c,v}
\cos^2\theta(\mb{k})
\frac{\partial}{\partial E_{n\mb{k}} } f_{n\mb{k}}   \nonumber \\
&=
\frac{e^2 g}{4\pi} 
\frac{v_F^2}{\omega^2+2i \omega \gamma}
\int_0^\infty k dk 
\frac{1}{\hbar v_F} \frac{\partial}{\partial k} ( f_{c\mb{k}} - f_{v\mb{k}} )
\nonumber  \\  
&=
-\frac{e^2 g}{4\pi\hbar} 
\frac{v_F k_F}{\omega^2 (1+2i\gamma/\omega) } .
\end{align}

For the standard relaxation operator, the expression of the susceptibility is given in Eq.~\eqref{eq_60_}. The contribution of the interband transitions is found to be
\begin{align}
\chi^{\mathrm{(st)}}_{\mathrm{inter}}(\omega, 0) 
&= 
- \frac{e^2 g}{4\pi^2} \int d^2 \mb{k} \left[ \frac{\sin^2\theta(\mb{k})}{4k^2} (f_{c\mb{k}}-f_{v\mb{k}}) \left( \frac{1}{2\hbar v_F k - \hbar\omega -i\hbar\gamma} - \frac{1}{-2\hbar v_F k - \hbar\omega -i\hbar\gamma} \right) \right] \nonumber \\
&= 
- \frac{e^2 g}{4\pi^2} \int d^2 \mb{k} \left[ \frac{\sin^2\theta(\mb{k})}{4k^2} (f_{c\mb{k}}-f_{v\mb{k}}) \frac{4 \hbar v_F k}{(2\hbar v_F k)^2 - (\hbar\omega +i\hbar\gamma)^2} \right] \nonumber \\
&= 
\frac{e^2 g}{4\pi} \int_{k_F}^\infty dk \left[ \frac{\hbar v_F}{(2\hbar v_F k)^2 - (\hbar\omega +i\hbar\gamma)^2} \right]   \nonumber \\
&=
-\frac{e^2 g}{16\pi}
\frac{1}{\hbar\omega (1+i\gamma/\omega) }
\ln
\left[
\frac{2 v_F k_F - \omega (1+i\gamma/\omega) } {2 v_F k_F + \omega (1+i\gamma/\omega) }
\right] .
\end{align}

The contribution of the intraband transitions is
\begin{align}  
{\chi}^{\mathrm{(st)}}_{\mathrm{intra}} \left(\omega, 0\right)
&=
\lim\limits_{\kappa \to 0}
\frac
{e^2g}{4\pi^2 \kappa^2}
 \int d^2 \mb{k} \sum_{n=c,v}
{\frac{ \kappa^2 \frac{1}{\hbar\omega + i\hbar\gamma} \left( \mb{n} \frac{\partial}{\partial \mb{k}} f_{n\mb{k}} \right) \left( \mb{n} \frac{\partial}{\partial \mb{k}} E_{n\mb{k}} \right) + \frac{1}{2} \kappa^2 \left( \mb{n} \frac{\partial}{\partial \mb{k}} \right)^2 f_{n\mb{k}} }
{\hbar\omega + i\hbar\gamma}}     \nonumber \\
&=
\frac
{e^2g}{4\pi^2} \frac{1}{\hbar\omega + i\hbar\gamma}
\int d^2 \mb{k} \sum_{n=c,v}
\left[
\frac{\left( \mb{n} \frac{\partial}{\partial \mb{k}} f_{n\mb{k}} \right) \left( \mb{n} \frac{\partial}{\partial \mb{k}} E_{n\mb{k}} \right)}{\hbar\omega + i\hbar\gamma} 
+ \frac{1}{2} \left( \mb{n} \frac{\partial}{\partial \mb{k}} \right)^2 f_{n\mb{k}}
\right] ,
\end{align} 
where the expansion in the numerator to the first order of $\kappa$ will disappear after integration. The derivative terms are 
\begin{align}
\left( \mb{n} \frac{\partial}{\partial \mb{k}} \right) E_{n\mb{k}} 
&=
s_n \hbar v_F \cos\theta(\mb{k})  ,  \nonumber \\
\left( \mb{n} \frac{\partial}{\partial \mb{k}} \right) f_{n\mb{k}} 
&=
\cos\theta(\mb{k}) \frac{\partial}{\partial k} f_{n\mb{k}} ,  \nonumber \\
\left( \mb{n} \frac{\partial}{\partial \mb{k}} \right)^2 f_{n\mb{k}}
&=
\cos^2\theta(\mb{k}) \frac{\partial^2}{\partial k^2} f_{n\mb{k}} + \frac{1}{k} \sin^2\theta(\mb{k}) \frac{\partial}{\partial k} f_{n\mb{k}} ,
\end{align}
where $s_n = 1,~-1$ for $n=c,~v$. 
The integration of the first term gives
\begin{align}
&\phantom{{}={}}
\int d^2 \mb{k} \sum_{n=c,v}
\left[ 
\cos^2\theta(\mb{k}) s_n \hbar v_F \frac{\partial}{\partial k} f_{n\mb{k}}
\right]    \nonumber \\
&=
\pi \sum_{n=c,v} s_n \hbar v_F \int_0^\infty k dk \frac{\partial}{\partial k} f_{n\mb{k}}
\nonumber \\
&=
-\pi \hbar v_F k_F .
\end{align}
The integration of the second term gives 
\begin{align}  
&\phantom{{}={}}
\frac{1}{2}
\int d^2 \mb{k} \sum_{n=c,v}
\left[ 
\cos^2\theta(\mb{k}) \frac{\partial^2}{\partial k^2} f_{n\mb{k}} + \frac{1}{k} \sin^2\theta(\mb{k}) \frac{\partial}{\partial k} f_{n\mb{k}}
\right]    \nonumber \\
&=
\frac{\pi}{2}
\int_0^\infty k dk \sum_{n=c,v}
\left[ 
\frac{\partial^2}{\partial k^2} f_{n\mb{k}} + \frac{1}{k} \frac{\partial}{\partial k} f_{n\mb{k}}
\right]  \nonumber \\
&=
\frac{\pi}{2}
\sum_{n=c,v} \int_0^\infty 
d \left( k \frac{\partial}{\partial k} f_{n\mb{k}} \right) 
 \nonumber \\
&=
0 .
\end{align} 
The result is  
\begin{align}
{\chi}^{\mathrm{(st)}}_{\mathrm{intra}} \left(\omega, 0\right)
&=
\frac{e^2g}{4\pi^2} \frac{1}{\hbar\omega + i\hbar\gamma} 
\frac{-\pi \hbar v_F k_F}{\hbar\omega + i\hbar\gamma}  \nonumber \\
&=
-\frac{e^2g}{4\pi\hbar} \frac{v_F k_F}{\omega^2(1 + i\gamma/\omega)^2} .
\end{align}

\bibliography{relaxation_ref}

\end{document}